## Original Paper

# On the Accuracy of Phase Extraction from a Known-Frequency Noisy Sinusoidal Signal

Emmanuel Dervieux[1,2*], Florian Tilquin[1], Alexis Bisiaux[1] and Wilfried Uhring[2]

[1] *Biosency, Cesson-Sévigné, France*
[2] *ICube, University of Strasbourg and CNRS, Strasbourg, France*

THIS IS AN AUTHOR'S PROOF WHICH HAS BEEN ACCEPTED FOR PUBLICATION AFTER PEER-REVIEW.

ABSTRACT

Accurate phase extraction from sinusoidal signals is a crucial task in various signal processing applications. While prior research predominantly addresses the case of asynchronous sampling with unknown signal frequency, this study focuses on the more specific situation where synchronous sampling is possible, and the signal's frequency is known. In this framework, a comprehensive analysis of phase estimation accuracy in the presence of both additive and phase noises is presented. A closed-form expression for the asymptotic Probability Density Function (PDF) of the resulting phase estimator is validated by simulations depicting Root Mean Square Error (RMSE) trends in different noise scenarios. This estimator is asymptotically efficient, converging rapidly to its Cramèr-Rao Lower Bound (CRLB). Three distinct RMSE behaviours were identified based on SNR, sample count ($N$), and noise level: *(i)* saturation towards a random guess at low Signal to Noise Ratio (SNR), *(ii)* linear decrease with the square roots of $N$ and SNR at moderate noise levels, and *(iii)* saturation at high SNR towards a noise floor dependent on the phase noise level. By quantifying the impact of sample count, additive noise, and phase noise on phase estimation accuracy, this work provides valuable insights for designing systems requiring precise phase extraction, such as phase-based fluorescence assays or system identification.

*Corresponding author: emmanuel.dervieux@biosency.com.

## 1 Introduction

Spectral estimation plays a critical role in signal processing by characterising a signal's spectral attributes, including amplitudes and phase shifts. This topic has gathered considerable research interest for decades due to its numerous applications in various fields, such as telecommunications, radar, seismology, and power grid analysis[1, 2, 3, 4]. In the general case, the frequencies of interest $f_i$ of the signal under study are *a priori* unknown. Thus, it is exceedingly unlikely that given a sampling frequency $f_s$ and a sampling length $N$, the numbers $f_i{\cdot}N/f_s$ are integers. This condition—known as "asynchronous sampling"—leads to the infamous picket fence and spectral leaking effects[5], which may be mitigated by an appropriate windowing function choice[6], the use of all-phase Discrete Fourier Transform (DFT)[7, 8], or both[9], for example.

There are certain cases, however, for which the signal under study is purely sinusoidal with a known frequency. This scenario arises when characterising linear systems, which may be fed a sinusoidal excitation signal of known frequency, amplitude, and phase, while recording their output. The analysis of the attenuation and phase shift induced by the system at hand can then yield useful information. For instance, in the context of frequency-based Dual Lifetime Referencing (f-DLR)[10], the phase shift between a fluorescence excitation signal of known frequency and the re-emitted one can be used to accurately measure the concentration of a variety of analytes[11, 12, 13, 14]. In this situation—known as "synchronous sampling"—the number of samples taken, as well as the sampling and excitation frequencies $f$ and $f_s$, can be chosen so that $f{\cdot}N/f_s$ is an integer, which suppresses the above-mentioned deleterious effects[5, 15].

Yet, as far as we are aware, no comprehensive study has been conducted to characterise the achievable accuracy of phase estimation in such a synchronous sampling scenario. In this paper, we present theoretical developments leading to a closed-form expression of the asymptotic Probability Density Function (PDF) of the phase estimate of a noisy sinusoidal signal in the presence of both phase and additive noises. The presented derivations are supported by simulations results, showing the resulting phase Root Mean Square Error (RMSE) at different noise levels. We then show that the derived phase estimator is asymptotically efficient with a fast convergence. Finally, we discuss its asymptotic behaviour in the case of very high or very low noise levels and sample numbers.

## 2 Problem Formulation

In the remainder of this document, the objective is always to retrieve the phase $\varphi$ of a real discrete signal of length $N$ defined as

$$s_n = A_s \cdot \cos\left(\frac{2 \cdot \pi \cdot f_0 \cdot n}{f_s} + \varphi + p_n\right) + x_n, \text{ with } \begin{cases} x_n \overset{iid}{\sim} \mathcal{N}(0, \sigma_x^2) \\ p_n \overset{iid}{\sim} \mathcal{N}(0, \sigma_p^2) \\ n \in [\![0; N-1]\!] \end{cases} \tag{1}$$

with $f_0$ the frequency of the signal itself, $f_s$ its sampling frequency—always chosen such that $f_s > 2 \cdot f_0$, the Nyquist frequency—and $A_s$ its amplitude. The $x_n$ and $p_n$ random variables—of variances $\sigma_x^2$ and $\sigma_p^2$—represent additive measurement noise and sampling-induced phase noise, respectively. Of note, it is also considered that $A_s$ and $\varphi$, though unknown, remain constant throughout the acquisition duration $N/f_s$. A representative illustration of the issue at hand, involving most of the parameters introduced above, may be seen in Figure 1.

Typically, in an f-DLR sensing scheme, $A_s$ and $\varphi$ would correspond to:

1. the intensity of the collected light: a function of the quantum yield of the involved fluorophores, of their concentrations, and of the illumination and light collection parameters, and

2. the phase shift: function of the ratio of the different fluorophores species, conveying the concentration of the analyte of interest.

Hence, it is of particular importance to accurately estimate $\varphi$, and to characterise the influence of $\sigma_p$, $\sigma_x$, and $N$ on its RMSE, since it will directly influence the reachable accuracy on the measurement of a given analyte's concentration.

In the remainder of this article, we adopt the following notations: $x_n$ refers to the n-th element of a given vector $X$, $\square^\intercal$ is the transpose operator, $\mathbb{0}_N$ and $\mathbb{1}_N$ stand for the zero and unit vector in $\mathbb{R}^N$, respectively, $\Re(z)$ and $\Im(z)$ stand for the real and imaginary parts of a given complex number $z$, while $|z|$ and $\arg(z)$ stand for its modulus and argument. $\triangleq$ means "per definition", $\overline{\square}$ is the complex conjugate operator, $\mathcal{N}$ and $\mathcal{CN}$ stand for the normal and complex normal distributions, respectively, and $x \perp\!\!\!\perp y$ denotes the independence between two random variable $x$ and $y$. Finally, the Signal to Noise Ratio (SNR) of the measurement is defined as

$$\mathrm{SNR} = \frac{A_s^2}{2\cdot\sigma_x^2} \quad \text{and} \quad \mathrm{SNR_{dB}} = 10\cdot \log_{10}(\mathrm{SNR}) \tag{2}$$

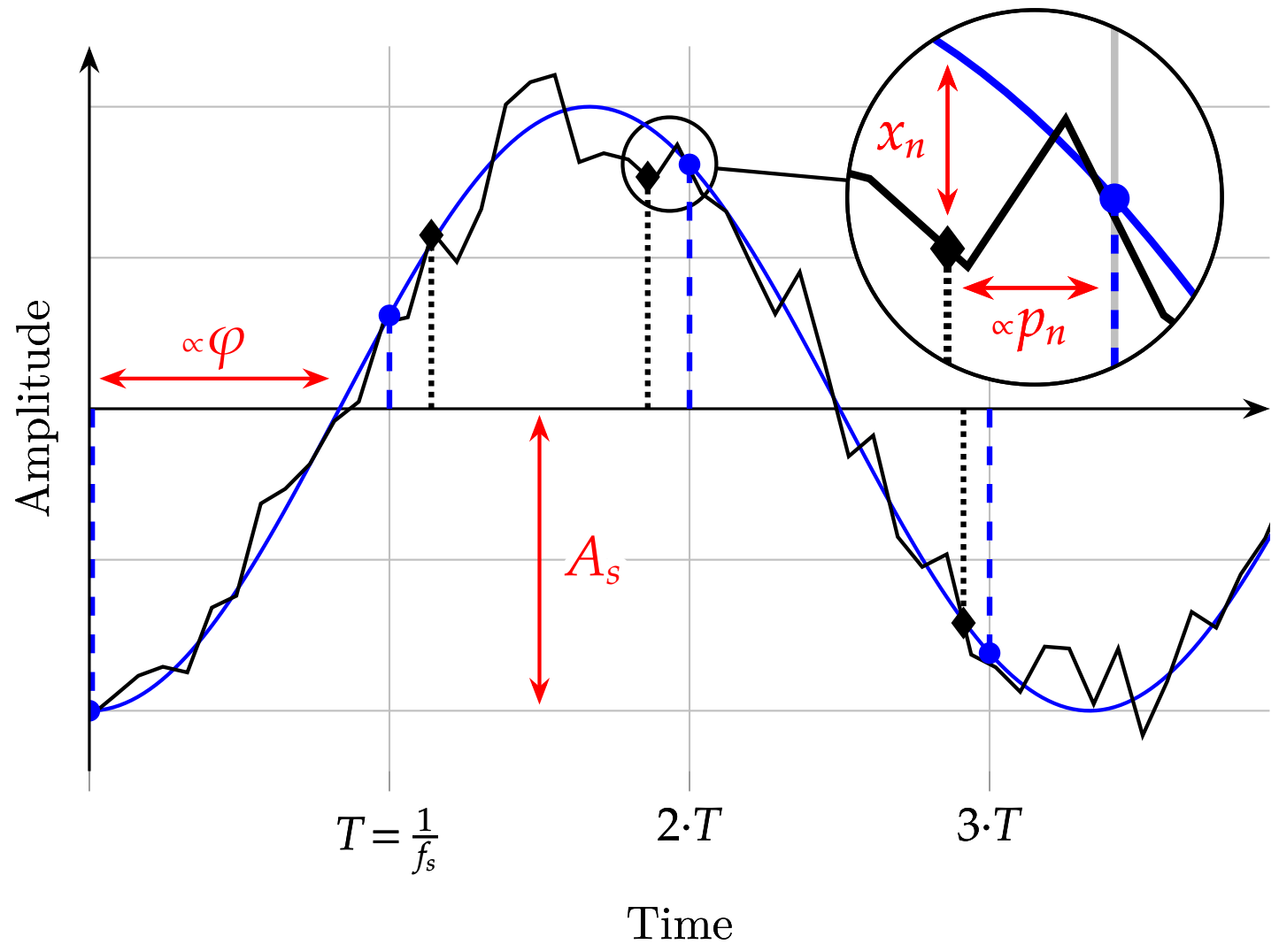

Figure 1: Temporal representation of the problem: the objective is to estimate the phase $\varphi$ of an idealised sinusoidal signal of amplitude $A_s$ (-●-) from a noisy measurement of the latter (-◆-). The noise originates from two distinct sources: *(i)* an additive noise $x_n$ which makes the sampled signal depart from its ideal counterpart, and *(ii)* a phase noise $p_n$, which randomly shifts the sampling times (▪▪▪) in comparison to an ideal sampling at frequency $f_s$ (- -). The symbol $\propto$ denotes proportionality, not to be confused with the Greek letter *alpha* ($\alpha$), used later on in this paper.

## 3 Characterisation of the DFT Distribution

This paper focuses on $\varphi$ estimation through the study of the DFT of the above-presented noisy signal. Indeed, we demonstrate in Section 4 that an unbiased and efficient estimator of $\varphi$—denoted as $\widehat{\varphi}$—can be derived by taking the argument of the signal's DFT at frequency $f_0$. In order to derive the PDF of $\widehat{\varphi}$, the PDF of this DFT must thus be known first.

To this end, let us first consider the $k$-th index of the $N$-points DFT of the above-mentioned signal[16], *i.e.* its DFT at frequency $f_0$:

$$D_N = \sum_{n=0}^{N-1} s_n \cdot \mathrm{e}^{-i\cdot\frac{2\cdot\pi\cdot k\cdot n}{N}} \tag{3}$$

with $k$, $N$, $f_0$ and $f_s$ chosen such that $\frac{k}{N} = \frac{f_0}{f_s}$ (synchronous sampling hypothesis). Let $\alpha = \frac{2\cdot\pi\cdot f_0}{f_s}$ and

$$S = \begin{pmatrix} s_0 \\ s_1 \\ \vdots \\ s_{N-1} \end{pmatrix},\ X = \begin{pmatrix} x_0 \\ x_1 \\ \vdots \\ x_{N-1} \end{pmatrix},\ Z = \begin{pmatrix} \mathrm{e}^{i\cdot p_0} \\ \mathrm{e}^{i\cdot p_1} \\ \vdots \\ \mathrm{e}^{i\cdot p_{N-1}} \end{pmatrix},\ \mathcal{U}_\alpha = \begin{pmatrix} 1 \\ \mathrm{e}^{-i\cdot\alpha} \\ \vdots \\ \mathrm{e}^{-i\cdot\alpha\cdot(N-1)} \end{pmatrix} \tag{4}$$

Then $D_N$ may be rewritten as

$$D_N = \sum_{n=0}^{N-1} s_n \cdot \mathrm{e}^{-i\cdot\alpha\cdot n} = \mathcal{U}_\alpha^\intercal \cdot S \tag{5}$$

Let also define $\widetilde{S}$ as $S = \widetilde{S} + X$. Using Euler's formula, each element $\widetilde{s}_n$ of $\widetilde{S}$ may then be expressed as

$$\begin{aligned} \widetilde{s}_n &= A_s \cdot \frac{\mathrm{e}^{i\cdot\left(\frac{2\cdot\pi\cdot f_0\cdot n}{f_s}+\varphi+p_n\right)} + \mathrm{e}^{-i\cdot\left(\frac{2\cdot\pi\cdot f_0\cdot n}{f_s}+\varphi+p_n\right)}}{2} \\ &= \frac{A_s}{2} \cdot \left(\mathrm{e}^{i\cdot(\alpha\cdot n+\varphi+p_n)} + \mathrm{e}^{-i\cdot(\alpha\cdot n+\varphi+p_n)}\right) \end{aligned} \tag{6}$$

The remainder of this section is organised as follows: the expected value and variance of $D_N$ are computed in Sections 3.1 and 3.2, respectively, while its asymptotic PDF is derived in Section 3.3.

### *3.1 Expected Value of $D_N$*

The expected value of $D_N$ is given by

$$\mathbf{E}[D_N] = \mathbf{E}[\mathcal{U}_\alpha^\intercal \cdot \widetilde{S}] + \mathbf{E}\left[\mathcal{U}_\alpha^\intercal \cdot X\right] \tag{7}$$

Regarding $\mathbf{E}[\mathcal{U}_\alpha^\intercal \cdot \widetilde{S}]$,

$$\begin{aligned} \mathcal{U}_\alpha^\intercal \cdot \widetilde{S} &= \frac{A_s}{2}\left(\mathrm{e}^{i\cdot\varphi} \sum_{n=0}^{N-1} \mathrm{e}^{i\cdot p_n} + \mathrm{e}^{-i\cdot\varphi} \cdot \sum_{n=0}^{N-1} \mathrm{e}^{-2\cdot i\cdot\alpha\cdot n} \cdot \mathrm{e}^{-i\cdot p_n}\right) \\ &= \frac{A_s}{2} \cdot \left(\mathrm{e}^{i\cdot\varphi} \cdot \mathbb{1}_N^\intercal \cdot Z + \mathrm{e}^{-i\cdot\varphi} \cdot \left(\mathcal{U}_\alpha^2\right)^\intercal \cdot \overline{Z}\right) \end{aligned} \tag{8}$$

hence

$$\mathbf{E}[\mathcal{U}_\alpha^\intercal \cdot \widetilde{S}] = \frac{A_s}{2} \cdot \left( \mathrm{e}^{i \cdot \varphi} \cdot \mathbb{1}_N^\intercal \cdot \mathbf{E}[Z] + \mathrm{e}^{-i \cdot \varphi} \cdot \left(\mathcal{U}_\alpha^2\right)^\intercal \cdot \mathbf{E}[\overline{Z}] \right) \tag{9}$$

We thus need to compute $\mathbf{E}[Z]$. For a given $p$ amongst $p_n$, let us consider $\mathbf{E}[\mathrm{e}^{i \cdot p}]$ first, and let $\left[g : v \mapsto \mathrm{e}^{i \cdot v}\right]$. Then

$$\mathbf{E}[\mathrm{e}^{i \cdot p}] = \mathbf{E}[g(p)] = \int_{-\infty}^{+\infty} g(v) \cdot f_p(v) dv \tag{10}$$

thanks to the Law Of The Unconscious Statistician (LOTUS)[17], wherein $f_p$ is the probability density function of $p$—namely a centred normal distribution of variance $\sigma_p^2$, see Equation 1—given by

$$\left[ f_p : v \mapsto \frac{1}{\sigma_p \cdot \sqrt{2 \cdot \pi}} \cdot \mathrm{e}^{-\frac{v^2}{2 \cdot \sigma_p^2}} \right] \tag{11}$$

We thus have

$$\begin{aligned} \mathbf{E}[\mathrm{e}^{i \cdot p}] &= \int_{-\infty}^{+\infty} \frac{1}{\sigma_p \cdot \sqrt{2 \cdot \pi}} \cdot \mathrm{e}^{-\frac{v^2}{2 \cdot \sigma_p^2} + i \cdot v} dv \\ &= \frac{\mathrm{e}^{-\frac{\sigma_p^2}{2}}}{\sigma_p \cdot \sqrt{2 \cdot \pi}} \int_{-\infty}^{+\infty} \mathrm{e}^{\left(\frac{i \cdot v}{\sigma_p \cdot \sqrt{2}} + \frac{\sigma_p}{\sqrt{2}}\right)^2} dv \end{aligned} \tag{12}$$

Using the change of variable $u = h(v)$ with

$$\left[ h : v \mapsto \frac{i \cdot v}{\sigma_p \cdot \sqrt{2}} + \frac{\sigma_p}{\sqrt{2}} \right] \tag{13}$$

we thus have $du = \frac{i \cdot dv}{\sigma_p \cdot \sqrt{2}}$, with erfi being the imaginary error function,

$$\begin{aligned} \mathbf{E}[\mathrm{e}^{i \cdot p}] &= \frac{\mathrm{e}^{-\frac{\sigma_p^2}{2}}}{\sigma_p \cdot \sqrt{2 \cdot \pi}} \cdot \lim_{v \to +\infty} \int_{-h(v)}^{+h(v)} \mathrm{e}^{u^2} \cdot \frac{\sigma_p \cdot \sqrt{2}}{i} du \\ &= \frac{\mathrm{e}^{-\frac{\sigma_p^2}{2}}}{2 \cdot i} \cdot \underbrace{\lim_{v \to +\infty} \underbrace{\frac{2}{\sqrt{\pi}} \int_{-h(v)}^{+h(v)} \mathrm{e}^{u^2} du}_{= [\operatorname{erfi}(u)]_{-h(v)}^{+h(v)}}}_{= 2 \cdot i} = \mathrm{e}^{-\frac{\sigma_p^2}{2}} \end{aligned} \tag{14}$$

Back to $\mathbf{E}[\mathcal{U}_\alpha^\intercal \cdot \widetilde{S}]$, since

$$\mathbf{E}[\overline{Z}] = \overline{\mathbf{E}[Z]} = \mathbf{E}[Z] = \mathbb{1}_N \cdot \mathrm{e}^{-\frac{\sigma_p^2}{2}} \tag{15}$$

we thus have

$$\mathbf{E}[\mathcal{U}_\alpha^\intercal \cdot \widetilde{S}] = \frac{A_s}{2} \cdot \left( \mathrm{e}^{i \cdot \varphi} \cdot N \cdot \mathrm{e}^{-\frac{\sigma_p^2}{2}} + \mathrm{e}^{-i \cdot \varphi} \cdot \mathrm{e}^{-\frac{\sigma_p^2}{2}} \cdot \sum_{n=0}^{N-1} u_{\alpha,n}^2 \right) \tag{16}$$

Noting that $\sum_{n=0}^{N-1} u_{\alpha,n}^2 = 0$ because $\mathrm{e}^{-2 \cdot i \cdot \alpha}$ is an N-th root of unity, comes

$$\mathbf{E}[\mathcal{U}_\alpha^\intercal \cdot \widetilde{S}] = \frac{A_s \cdot N}{2} \cdot \mathrm{e}^{i \cdot \varphi} \cdot \mathrm{e}^{-\frac{\sigma_p^2}{2}} \tag{17}$$

Regarding $\mathbf{E}[\mathcal{U}_\alpha^\intercal \cdot X]$,

$$\mathbf{E}[\mathcal{U}_\alpha^\intercal \cdot X] = \mathcal{U}_\alpha^\intercal \cdot \underbrace{\mathbf{E}[X]}_{=\mathbb{0}_N} = 0 \tag{18}$$

Finally, the expected value of $D_N$ comes to be

$$\boxed{\mathbf{E}[D_N] = \frac{A_s \cdot N}{2} \cdot \mathrm{e}^{i \cdot \varphi} \cdot \mathrm{e}^{-\frac{\sigma_p^2}{2}}} \tag{19}$$

### *3.2 Variance of $D_N$*

The variance of $D_N$ may also be calculated in a similar manner, starting with

$$\mathbf{Var}(D_N) = \mathbf{Var}(\mathcal{U}_\alpha^\intercal \cdot \widetilde{S}) + \mathbf{Var}\left(\mathcal{U}_\alpha^\intercal \cdot X\right) \tag{20}$$

Regarding $\mathbf{Var}(\mathcal{U}_\alpha^\intercal \cdot \widetilde{S})$,

$$\begin{aligned} \mathbf{Var}(\mathcal{U}_\alpha^\intercal \cdot \widetilde{S}) &= \mathbf{Var}\left[ \frac{A_s}{2} \cdot \left( \mathrm{e}^{i \cdot \varphi} \cdot \mathbb{1}_N^\intercal \cdot Z + \mathrm{e}^{-i \cdot \varphi} \cdot \left(\mathcal{U}_\alpha^2\right)^\intercal \cdot \overline{Z} \right) \right] \\ &= \frac{A_s^2}{4} \cdot \left( \sum_{n=0}^{N-1} \mathbf{Var}(z_n) + \mathbf{Var}\left(\overline{z_n}\right) \right) \end{aligned} \tag{21}$$

We thus need to compute $\mathbf{Var}(z_n)$, *i.e.* $\mathbf{Var}(\mathrm{e}^{i \cdot p_n})$.

$$\begin{aligned} \mathbf{Var}(\mathrm{e}^{i \cdot p}) &= \mathbf{E}\left[ \left| \mathrm{e}^{i \cdot p} \right|^2 \right] - \left| \mathbf{E}\left[ \mathrm{e}^{i \cdot p} \right] \right|^2 \\ &= 1 - \mathrm{e}^{-\sigma_p^2} = \mathbf{Var}(\mathrm{e}^{-i \cdot p}) \end{aligned} \tag{22}$$

as $\mathbf{E}[e^{-i \cdot p}] = \overline{\mathbf{E}[e^{i \cdot p}]}$. Thus

$$\mathbf{Var}(\mathcal{U}_\alpha^\intercal \cdot \widetilde{S}) = \frac{N \cdot A_s^2}{2} \cdot \left( 1 - \mathrm{e}^{-\sigma_p^2} \right) \tag{23}$$

Regarding $\mathbf{Var}\left(\mathcal{U}_\alpha^\intercal \cdot X\right)$,

$$\begin{aligned}\mathbf{Var}\left(\mathcal{U}_\alpha^\intercal \cdot X\right) &= \sum_{n=0}^{N-1} \mathbf{Var}\left(u_{\alpha,n} x_n\right) = \sum_{n=0}^{N-1} \left(\mathbf{E}\left[\left|u_{\alpha,n} x_n\right|^2\right] - \left|\mathbf{E}\left[u_{\alpha,n} x_n\right]\right|^2\right) \\ &= \sum_{n=0}^{N-1} \mathbf{E}[x_n^2] = N \cdot \sigma_x^2\end{aligned} \tag{24}$$

Finally, the variance of $D_N$ comes to be

$$\boxed{\mathbf{Var}(D_N) = N \cdot \left(\frac{A_s^2}{2} \cdot \left(1 - \mathrm{e}^{-\sigma_p^2}\right) + \sigma_x^2\right)} \tag{25}$$

### 3.3 Distribution of $D_N$

Now that the expected value and variance of $D_N$ are known, the next step is to study its PDF. To do so, we focus on a reduced version of $D_N$—$\widetilde{D}_N$—defined as $\widetilde{D}_N = \frac{2 \cdot D_N}{A_s \cdot N}$, with

$$\beta_p = \mathrm{e}^{-\frac{\sigma_p^2}{2}} \qquad \begin{aligned} \mathbf{E}[\widetilde{D}_N] &= \beta_p \cdot \mathrm{e}^{i \cdot \varphi} \\ \mathbf{Var}(\widetilde{D}_N) &= \tfrac{2}{N} \cdot \left(1 - \beta_p^2 + \tfrac{1}{\mathrm{SNR}}\right) \end{aligned} \tag{26}$$

We then proceed in two steps: at first, the convergence in law of $\Re(\widetilde{D}_N)$ and $\Im(\widetilde{D}_N)$ towards normal distributions is demonstrated. Then, the asymptotic independence of the latter two quantities is shown. These two demonstrations establish that $\widetilde{D}_N$ converges in law[1] towards a complex normal distribution[19, pp. 540–559], which is a crucial requirement for the forthcoming developments (see Section 4). However, before delving any deeper into this two-step demonstration, we can further simplify the issue at hand, observing that

$$\widetilde{D}_N = \underbrace{\frac{2 \cdot \mathcal{U}_\alpha^\intercal \cdot \widetilde{S}}{A_s \cdot N}}_{\widetilde{D}_s} + \underbrace{\frac{2 \cdot \mathcal{U}_\alpha^\intercal \cdot X}{A_s \cdot N}}_{\widetilde{D}_x} \tag{27}$$

Since $X$ is stationary and ergodic, it readily follows that $\widetilde{D}_x$ converges in distribution toward a complex normal distribution[20, 21]. Since $\widetilde{D}_s$ and $\widetilde{D}_x$ are independent, the two above-mentioned steps thus only have to be performed for $\widetilde{D}_s$.

[1] Of note, convergence in law is sometimes also referred to as "convergence in distribution" or "weak convergence"[18, p. 18].

### 3.3.1 Convergence in Law Towards a Normal Distribution

Let us consider the real part of $\widetilde{D}_s$

$$\begin{aligned}\Re(\widetilde{D}_s) &= \frac{2}{N}\cdot\sum_{n=0}^{N-1}\cos(\alpha\cdot n)\cdot\cos(\alpha\cdot n+\varphi+p_n)\\ &= \frac{2}{N}\cdot\sum_{n=0}^{N-1}V_n,\text{ with }V_n=\cos(\alpha\cdot n)\cdot\cos(\alpha\cdot n+\varphi+p_n)\end{aligned}\tag{28}$$

We will use Lyapunov's Central Limit Theorem (L-CLT) to demonstrate the convergence of $\sum V_n$ towards a normal distribution. To do so, we will first show that there exists a positive $\delta$ such that

$$\lim_{N\to+\infty}\underbrace{\frac{1}{s_N^{2+\delta}}\sum_{n=0}^{N-1}\mathbf{E}\left[\left|V_n-\mathbf{E}[V_n]\right|^{2+\delta}\right]}_{\gamma_N}=0\tag{29}$$

wherein $s_{\mathrm{N}}^2=\sum_{n=0}^{N-1}\mathbf{Var}(V_n)$, and $\mathbf{Var}(V_n)=\cos^2(\alpha\cdot n)\cdot\underbrace{\mathbf{Var}(\cos(t_{\mathrm{n}}+p_n))}_{=\mathbf{E}(\ldots^2)-\mathbf{E}(\ldots)^2}$, with $t_n=\alpha\cdot n+\varphi$. Where

$$\mathbf{E}(\ldots^2)=\frac{1}{2}+\frac{e^{-2\cdot\sigma_p^2}}{2}\cdot\cos(2\cdot t_n),\text{ and }\mathbf{E}(\ldots)^2=e^{-\sigma_p^2}\cdot\frac{1+\cos(2\cdot t_n)}{2}\tag{30}$$

and thus

$$\begin{aligned}\mathbf{Var}(V_n) &= \cos^2(\alpha\cdot n)\cdot\left[\frac{1}{2}+e^{-\sigma_p^2}\cdot\left(\frac{\cos(2\cdot t_n)}{2}\cdot e^{-\sigma_p^2}-\frac{1+\cos(2\cdot t_n)}{2}\right)\right]\\ &= \frac{\cos^2(\alpha\cdot n)}{2}\cdot\left(1-e^{-\sigma_p^2}\right)\cdot\left(1-e^{-\sigma_p^2}\cdot\cos(2\cdot t_n)\right)\\ &\geq \frac{\cos^2(\alpha\cdot n)}{2}\cdot\left(1-e^{-\sigma_p^2}\right)^2\end{aligned}\tag{31}$$

Back to $s_n$,

$$s_n^2\geq\frac{\left(1-e^{-\sigma_p^2}\right)^2}{2}\overbrace{\sum_{n=0}^{N-1}\cos^2(\alpha\cdot n)}^{=N/2^{(\ddagger)}}\geq\frac{N\cdot\left(1-e^{-\sigma_p^2}\right)^2}{4}\tag{32}$$

wherein ($\ddagger$) comes from the facts that $\cos^2 x=\frac{1+\cos(2\cdot x)}{2}$, and that $e^{-2\cdot i\cdot\alpha}$ is an N-th root of unity (see Equations 16–17). Then, $\forall \delta > 0$

$$\gamma_N = \frac{1}{s_N^{2+\delta}} \sum_{n=0}^{N-1} \overbrace{\mathbf{E}\left[\left|V_n - \mathbf{E}[V_n]\right|^{2+\delta}\right]}^{\leq 2^{2+\delta}} \leq \frac{N \cdot 2^{2+\delta}}{\left(\frac{N}{4} \cdot \left(1 - e^{-\sigma_p^2}\right)^2\right)^{1+\frac{\delta}{2}}} \xrightarrow[N \to +\infty]{} 0 \tag{33}$$

Since $V_n$ are independent and of finite variance, according to L-CLT, we thus have

$$\frac{1}{s_N} \sum_{n=0}^{N-1} (V_n - \mathbf{E}[V_n]) \xrightarrow[N \to +\infty]{d} \mathcal{N}(0, 1) \tag{34}$$

wherein $\xrightarrow{d}$ denotes convergence in distribution.

Hence, since $\Re(\widetilde{D}_s) = \frac{2}{N} \cdot \sum_{n=0}^{N-1} V_n$,

$$\Re(\widetilde{D}_s) \xrightarrow[N \to +\infty]{d} \mathcal{N}\left(\frac{2}{N} \cdot \sum_{n=0}^{N-1} \mathbf{E}[V_n], \left(\frac{2 \cdot s_N}{N}\right)^2\right) \tag{35}$$

A similar train of thought can be followed to also demonstrate the asymptotic normality of $\Im(\widetilde{D}_s)$.

### *3.3.2 Asymptotic Independence*

Demonstrating the complex normality of $\widetilde{D}_s$ then only requires to demonstrate that $\Re(\widetilde{D}_s) \perp\!\!\!\perp \Im(\widetilde{D}_s)$. To do so, it suffices to show that[22, Th. 4.5-1]:

*(i)* $\Re(\widetilde{D}_s)$ and $\Im(\widetilde{D}_s)$ follow a bivariate normal distribution, and that

*(ii)* $\mathbf{Cov}(\Re(\widetilde{D}_s), \Im(\widetilde{D}_s)) = 0$.

**Bivariate normality**

$\forall (a, b) \in \mathbb{R}^2$ let

$$\begin{aligned} T &= a \cdot \Re(\widetilde{D}_s) + b \cdot \Im(\widetilde{D}_s) \\ &= \sum_{n=0}^{N-1} \frac{a \cdot \cos(\alpha \cdot n) + b \cdot \sin(\alpha \cdot n)}{N} \cdot \cos(\alpha \cdot n + \varphi + p_n) \end{aligned} \tag{36}$$

It can be shown—as was done in the previous section with $\Re(\widetilde{D}_s)$—that $T$ also converges in law towards a normal distribution. Thus, by definition, $\Re(\widetilde{D}_s)$ and $\Im(\widetilde{D}_s)$ follow a bivariate normal distribution.

**Covariance**

Let

$$C = \mathbf{Cov}(\Re(\widetilde{D}_s), \Im(\widetilde{D}_s)) = \underbrace{\mathbf{E}[\Re(\widetilde{D}_s)\cdot\Im(\widetilde{D}_s)]}_{C_L} - \underbrace{\mathbf{E}[\Re(\widetilde{D}_s)]\cdot\mathbf{E}[\Im(\widetilde{D}_s)]}_{C_R} \tag{37}$$

and $\forall n$, $\Phi_n = \varphi + p_n$. Then

$$\begin{aligned} C_L &= \mathbf{E}\left[\left(\sum_{l=0}^{N-1} \frac{\cos(\alpha\cdot l)\cdot\cos(\alpha\cdot l + \Phi_l)}{N}\right)\cdot\left(\sum_{l=0}^{N-1} \frac{\sin(\alpha\cdot m)\cdot\cos(\alpha\cdot m + \Phi_m)}{N}\right)\right] \\ &= \sum_{l,m=0}^{N-1}\left(\frac{\cos(\alpha\cdot l)\cdot\sin(\alpha\cdot m)}{N^2}\cdot\mathbf{E}\left[\cos(\alpha\cdot l + \Phi_l)\cdot\cos(\alpha\cdot m + \Phi_m)\right]\right) \end{aligned} \tag{38}$$

while

$$C_R = \sum_{l,m=0}^{N-1}\left(\frac{\cos(\alpha\cdot l)\cdot\sin(\alpha\cdot m)}{N^2}\cdot\mathbf{E}\left[\cos(\alpha\cdot l + \Phi_l)\right]\cdot\mathbf{E}\left[\cos(\alpha\cdot m + \Phi_m)\right]\right) \tag{39}$$

Since $(p_n)$ are independent, $(\Phi_n)$ are also independent and $\forall l \neq m$

$$\mathbf{E}\left[\cos(\alpha\cdot l + \Phi_l)\cdot\cos(\alpha\cdot m + \Phi_m)\right] = \mathbf{E}\left[\cos(\alpha\cdot l + \Phi_l)\right]\cdot\mathbf{E}\left[\cos(\alpha\cdot m + \Phi_m)\right] \tag{40}$$

The corresponding terms in $C_L$ and $C_R$ thus cancel out and

$$C = \frac{1}{N^2}\sum_{n=0}^{N-1}\cos(\alpha\cdot n)\cdot\sin(\alpha\cdot n)\cdot\underbrace{\left[\mathbf{E}\left[\cos^2(\alpha\cdot n + \Phi_n)\right] - \mathbf{E}\left[\cos(\alpha\cdot n + \Phi_n)\right]^2\right]}_{=\mathbf{Var}(\cos(\alpha\cdot n + \Phi_n))} \tag{41}$$

Using Popoviciu's inequality on variances to bound $\mathbf{Var}(\cos(...))$ yields

$$C \leq \frac{1}{N^2}\sum_{n=0}^{N-1}\underbrace{|\cos(\alpha\cdot n)|}_{\leq 1}\cdot\underbrace{|\sin(\alpha\cdot n)|}_{\leq 1}\cdot\underbrace{\mathbf{Var}(\cos(...))}_{\leq 1} \leq \frac{1}{N}\xrightarrow[N\to+\infty]{} 0 \tag{42}$$

Thus $\Re(\widetilde{D}_s) \perp\!\!\!\perp \Im(\widetilde{D}_s)$ asymptotically, and we finally demonstrated that $\widetilde{D}_s$—and thus $\widetilde{D}_N$—converges in law towards a complex normal distribution with increasing values of $N$. This can be rephrased as

$$\begin{cases} \widetilde{D}_N = x + i\cdot y \\ x \sim \mathcal{N}\left(\mu_x = \beta_p\cdot\cos(\varphi), \sigma^2\right) \\ y \sim \mathcal{N}\left(\mu_y = \beta_p\cdot\sin(\varphi), \sigma^2\right) \\ x \perp\!\!\!\perp y \end{cases} \quad \text{wherein } \sigma^2 = \frac{1}{N}\cdot\left(1 - \beta_p^2 + \frac{1}{\text{SNR}}\right) \tag{43}$$

### 3.3.3 Convergence in Practice

While this convergence is theoretically proven for $N \to +\infty$, its practical validity was studied for relatively small values of $N$. To do so, simulations were performed for $\text{SNR}_{\text{dB}}$= 0..30 dB, $\sigma_p = 0.1..10°$, $N = 10..100$. The multivariate normality of $\Re(\widetilde{D}_N)$ and $\Im(\widetilde{D}_N)$ was tested using Henze-Zirkler test[23] on $2.10^3$ $\widetilde{D}_N$ outcomes. Their independence, was asserted using Hoeffding's D measures[24] between $\Re(\widetilde{D}_N)$ and $\Im(\widetilde{D}_N)$ using $10^5$ $\widetilde{D}_N$ outcomes. In the case of the Henze-Zirkler test, simulations were repeated ten times, and their p-values—adjusted using the Benjamini-Hochberg correction[25] and combined with Fisher's method. The outcomes of these simulations were as follows. Henze-Zirkler multivariate test revealed no deviation from normality for $N$ values above 20 using a significance level of 0.05. Hoeffding's D measures between $\Re(\widetilde{D}_N)$ and $\Im(\widetilde{D}_N)$, on their part, were found to be below $10^{-4}$ for as little as $N = 20$ measurement points, clearly demonstrating the practical independence of the two random variables. These two tests confirm that the convergence of $\widetilde{D}_N)$ towards a complex normal distribution is observed in practice across a wide range of conditions, even for relatively small $N$ values. Of note, this fast convergence is again demonstrated in Figure 3, further below, for as little as 20 samples.

## 4 Phase Estimation from the DFT

The phase $\varphi$ of the signal $S$ may be estimated by $\widehat{\varphi} = \arg(\widetilde{D}_N)$. Of paramount importance are thus the mean bias and RMSE of this estimator, *i.e.* the two quantities

$$\mathbf{E}[\varphi - \widehat{\varphi}] \quad \text{and} \quad \sqrt{\mathbf{E}\left[(\varphi - \widehat{\varphi})^2\right]} \tag{44}$$

In order to derive them, we can notice that $\widehat{\varphi} = \arg(x + i \cdot y)$, and focus on the joint probability function of $(x, y)$, given by

$$f(x, y) = \frac{1}{2 \cdot \pi \cdot \sigma^2} \cdot \mathrm{e}^{-\frac{1}{2 \cdot \sigma^2} \cdot ((x-\mu_x)^2 + (y-\mu_y)^2)} \tag{45}$$

Then—thanks to the LOTUS—$\widehat{\varphi}$'s mean bias ($a = 1$) and RMSE ($a = 2$) may be computed using

$$\mathbf{E}\left[(\varphi - \widehat{\varphi})^a\right] = \iint_{\mathbb{R}^2} \arg\left(x + i \cdot y\right)^a \cdot f(x, y)\; dxdy \tag{46}$$

While direct numerical calculations are presented in Section 4.3, another approach involving a switch to polar coordinates is first presented in the next section, allowing an in-depth comprehension of the influence of noises by means of meaningful illustrations.

### *4.1 Marginalisation in Polar Coordinates*

A representation of the estimation of $\varphi$ in the complex plane can be seen in Figure 2. Naturally, different realisations of $S$ would lead to different $\widetilde{D}_N$ values. Since $\widetilde{D}_N$ follows a complex normal distribution, this translates into the small black dots ($\cdot$) whose repartition is characteristic of a bivariate normal law, in the complex plane. This distribution is centred around its mean—$\mathbf{E}[\widetilde{D}_N]$—of Cartesian coordinates $(\mu_x, \mu_y)_c$, represented as a large black dot ($\bullet$). Interestingly, this centre is distinct from its position in the noiseless case, represented as a large blue dot ($\bullet$), due to the $\mathrm{e}^{-\sigma_p^2}$ term in $\mathbf{E}[\widetilde{D}_N]$, which is dragging the $\widetilde{D}_N$ distribution towards the origin. A given $\widetilde{D}_N$ realisation—depicted as a large red dot ($\bullet$)—yields an estimation of $\varphi$, namely $\widehat{\varphi}$.

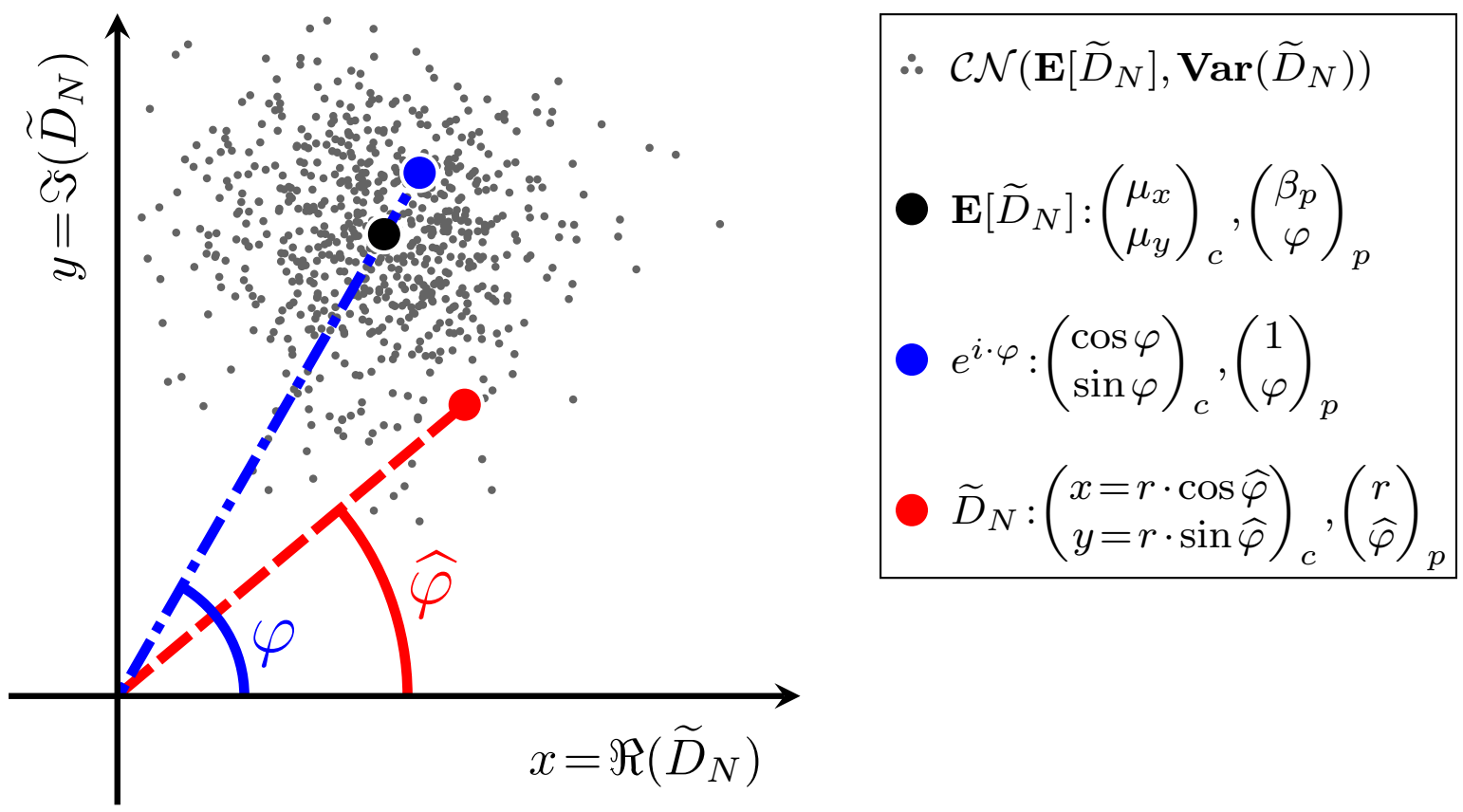

Figure 2: A representation of the estimation issue at hand in the complex plane. The $(\square)_c$ and $(\square)_p$ subscripts denote Cartesian and polar coordinates, respectively. See the text for further explanations.

Formally, the following change of variable from Cartesian to polar coordinates may be performed:

$$\left[\mathbf{F} : \begin{array}{c} \mathbb{R}^+ \times [0; 2\cdot\pi[ \to \mathbb{R}^2 \\ \begin{pmatrix} r \\ \theta \end{pmatrix} \mapsto \begin{pmatrix} r\cdot \cos\ \theta \\ r\cdot \sin\ \theta \end{pmatrix} \end{array}\right] \text{ and } \quad \mathbf{J}_{\mathbf{F}}(r, \theta) = r \tag{47}$$

wherein $\mathbf{J_F}$ is the Jacobian of $\mathbf{F}$. Expressing $f$ as a function of $r$ and $\theta$ yields

$$\begin{aligned} f(r,\theta) &= \frac{1}{2\cdot\pi\cdot\sigma^2}\cdot \mathrm{e}^{-\frac{1}{2\cdot\sigma^2}\cdot\left(r^2-2\cdot r\cdot\beta_p\cdot\cos(\theta-\varphi)+\beta_p^2\right)} \\ &= \frac{1}{2\cdot\pi\cdot\sigma^2}\cdot \mathrm{e}^{-\frac{(r-\beta_p\cdot\cos(\theta-\varphi))^2+\beta_p^2\cdot\sin^2(\theta-\varphi)}{2\cdot\sigma^2}} \\ &= \frac{\mathrm{e}^{-\beta_p^2\cdot\frac{\sin^2(\theta-\varphi)}{2\cdot\sigma^2}}}{2\cdot\pi\cdot\sigma^2}\cdot \mathrm{e}^{-\frac{1}{2\cdot\sigma^2}\cdot(r-\beta_p\cdot\cos(\theta-\varphi))^2} \end{aligned} \tag{48}$$

Applying the above-mentioned change of variables, taking care to replace $dxdy$ by $\mathbf{J_F}(r,\theta)drd\theta$ in the integrand leads to the following expression for the marginal density probability of $\theta$ under $\varphi$:

$$\iint\limits_{(x,y)\in\mathbb{R}^2} f(x,y)dxdy = \iint\limits_{\substack{r\in\mathbb{R}^+ \\ \theta\in[0;2\cdot\pi[}} f(r,\theta)\cdot r\ drd\theta \tag{49}$$

and $f$ can be marginalised with respect to $r$ alone, leading to

$$g_\varphi(\theta) \triangleq \int_0^{+\infty} f(\theta,r)\cdot r\ dr = \frac{\mathrm{e}^{-\beta_p^2\cdot\frac{\sin^2(\theta-\varphi)}{2\cdot\sigma^2}}}{2\cdot\pi\cdot\sigma^2}\cdot \underbrace{\int_0^{+\infty} r\cdot\mathrm{e}^{-\frac{(r-\beta_p\cdot\cos(\theta-\varphi))^2}{2\cdot\sigma^2}}dr}_{\mathcal{G}_\theta} \tag{50}$$

$\mathcal{G}_\theta$ can be further decomposed into $\mathcal{H}_\theta$ and $\mathcal{K}_\theta$, following

$$\begin{aligned} \mathcal{G}_\theta = &\overbrace{\int_0^{+\infty} (r-\beta_p\cdot\cos(\theta-\varphi))\cdot\mathrm{e}^{-\frac{(r-\beta_p\cdot\cos(\theta-\varphi))^2}{2\cdot\sigma^2}}dr}^{\mathcal{H}_\theta} \\ &+\beta_p\cdot\cos(\theta-\varphi)\cdot\underbrace{\int_0^{+\infty} \mathrm{e}^{-\frac{1}{2\cdot\sigma^2}\cdot(r-\beta_p\cdot\cos(\theta-\varphi))^2}dr}_{\mathcal{K}_\theta} \end{aligned} \tag{51}$$

And $\mathcal{H}_\theta$ and $\mathcal{K}_\theta$ can be explicitly calculated using appropriate change of variables as

$$\mathcal{H}_\theta = \int_{-\beta_p\cdot\cos(\theta-\varphi)}^{+\infty} t\cdot\mathrm{e}^{-\frac{t^2}{2\cdot\sigma^2}}dt = -\sigma^2\cdot\int_{-\beta_p\cdot\cos(\theta-\varphi)}^{+\infty}\frac{d}{dt}\left(\mathrm{e}^{-\frac{t^2}{2\cdot\sigma^2}}\right)dt = \sigma^2\cdot\mathrm{e}^{-\beta_p^2\cdot\frac{\cos^2(\theta-\varphi)}{2\cdot\sigma^2}} \tag{52}$$

and

$$\mathcal{K}_\theta = \int_{-\beta_p\cdot\cos(\theta-\varphi)}^{+\infty} \mathrm{e}^{-\frac{t^2}{2\cdot\sigma^2}}\,dt = \int_{-\beta_p\cdot\cos(\theta-\varphi)}^{+\infty} \mathrm{e}^{-\left(\frac{t}{\sqrt{2}\cdot\sigma}\right)^2}\,dt = \sigma\cdot\sqrt{2}\cdot\int_{-\beta_p\cdot\frac{\cos(\theta-\varphi)}{\sigma\cdot\sqrt{2}}}^{+\infty} \mathrm{e}^{-u^2}\,du$$
$$= \sigma\cdot\sqrt{\frac{\pi}{2}}\cdot\underbrace{\frac{2}{\sqrt{\pi}}\cdot\int_{-\beta_p\cdot\frac{\cos(\theta-\varphi)}{\sigma\cdot\sqrt{2}}}^{+\infty} \mathrm{e}^{-u^2}\,du}_{\mathrm{erfc}(\ldots)} = \sigma\cdot\sqrt{\frac{\pi}{2}}\cdot\mathrm{erfc}\left(-\beta_p\cdot\frac{\cos(\theta-\varphi)}{\sigma\cdot\sqrt{2}}\right) \tag{53}$$

and thus:

$$\mathcal{G}_\theta = \sigma^2\cdot\mathrm{e}^{-\beta_p^2\frac{\cos^2(\theta-\varphi)}{2\cdot\sigma^2}} + \sigma\cdot\beta_p\cdot\sqrt{\frac{\pi}{2}}\cdot\cos(\theta-\varphi)\cdot\mathrm{erfc}\left(-\beta_p\cdot\frac{\cos(\theta-\varphi)}{\sigma\cdot\sqrt{2}}\right) \tag{54}$$

Finally, putting it all together leads to

$$\boxed{g_\varphi(\theta) = \frac{\mathrm{e}^{-\frac{\beta_p^2}{2\sigma^2}}}{2\cdot\pi} + \frac{\beta_p\cdot\cos(\theta-\varphi)\cdot\mathrm{e}^{-\beta_p^2\cdot\frac{\sin^2(\theta-\varphi)}{2\cdot\sigma^2}}}{2\cdot\sqrt{2\cdot\pi}\cdot\sigma}\cdot\mathrm{erfc}\left(-\beta_p\cdot\frac{\cos(\theta-\varphi)}{\sigma\cdot\sqrt{2}}\right)} \tag{55}$$

$g_\varphi(\theta)$ is represented in Figure 3, along with the histogram of simulated $\widehat{\varphi}$ values—*i.e.* $\arg(\widetilde{D}_N)$ values, $\widetilde{D}_N$ being computed from simulated $S$ vectors. $g_\varphi(\theta)$ resembles a normal distribution which would have been wrapped around the $[0, 2\cdot\pi]$ interval—although it is not a wrapped normal distribution nor a von Mises distribution strictly speaking (for further information on these distributions, see Collett, Mardia, Ley *et al.*[26, 27, 28]). As could have been expected intuitively, the shape of $g_\varphi(\theta)$ narrows as $N$ or $\mathrm{SNR_{dB}}$ increase, or as $\sigma_p$ decreases, as emphasised in the inset of Figure 3.

### *4.2 Expressing the Error in the Polar Case*

Since $g_\varphi(\theta)$ is the probability density function to measure a phase shift $\widehat{\varphi} = \theta$ given a true phase shift $\varphi$, the mean bias and RMSE of the $\widehat{\varphi}$ estimator, may be expressed as

$$\underset{\text{(bias)}}{\mathbf{E}[\varphi - \widehat{\varphi}]} = \int_0^{2\cdot\pi} err_\varphi(\theta)\cdot g_\varphi(\theta)d\theta, \text{ and}$$
$$\mathrm{RMSE} = \sqrt{\int_0^{2\cdot\pi} err_\varphi^2(\theta)\cdot g_\varphi(\theta)d\theta} \tag{56}$$

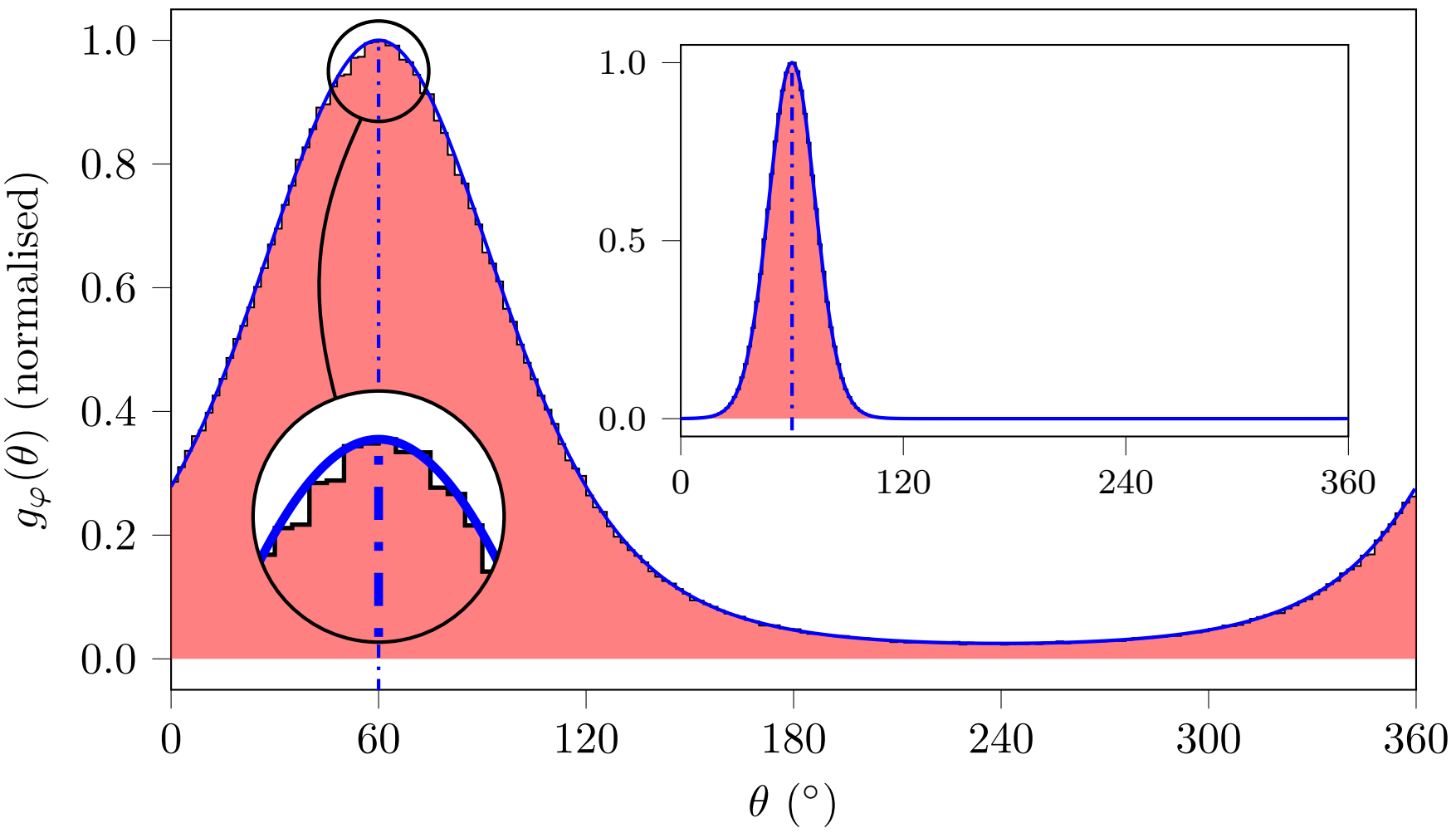

Figure 3: Normalised representations of $g_\varphi(\theta)$ and of the histogram of simulated $\widehat{\varphi}$ values ($10^6$ draws, $\varphi$=60°). **Main picture:** $N = 20$, $\text{SNR}_{\text{dB}}$=-10 dB and $\sigma_p = 5°$. The magnifying glass clearly shows how the histogram of simulated $\widehat{\varphi}$ is close to the analytic expression of $g_\varphi(\theta)$. **Inset:** $N = 20$, $\text{SNR}_{\text{dB}}$=0 dB and $\sigma_p = 1°$.

wherein $err_\varphi(\theta)$ is the estimation error at a given polar angle $\theta$, as represented in Figure 4, and is defined as

$$err_\varphi(\theta) = \min\left(d, 2{\cdot}\pi - d\right) \tag{57}$$

with $d \equiv \theta - \varphi \pmod{2{\cdot}\pi}$, that is

$$err_\varphi(\theta) = \pi - |\pi - \theta| \tag{58}$$

It can also be demonstrated that the RMSE and bias are independent of $\varphi$, as could have been expected intuitively. Only the RMSE case is detailed below, but a similar train of thought can be followed for the bias. Since the square root function is bijective on $\mathbb{R}^+$, we only have to demonstrate that $\forall(\varphi, \varphi') \in [0, 2{\cdot}\pi[^2$

$$\underbrace{\int_0^{2\cdot\pi} err_\varphi^2(\theta){\cdot}g_\varphi(\theta)d\theta}_{\mathcal{J}_\varphi} = \underbrace{\int_0^{2\cdot\pi} err_{\varphi'}^2(\theta){\cdot}g_{\varphi'}(\theta)d\theta}_{\mathcal{J}_{\varphi'}} \tag{59}$$

Then, with the change of variable $t = \theta - \varphi + \varphi'$,

$$\mathcal{J}_\varphi = \int_{-\varphi+\varphi'}^{2\cdot\pi-\varphi+\varphi'} err_\varphi^2(t + \varphi - \varphi'){\cdot}g_\varphi(t + \varphi - \varphi')d\theta \tag{60}$$

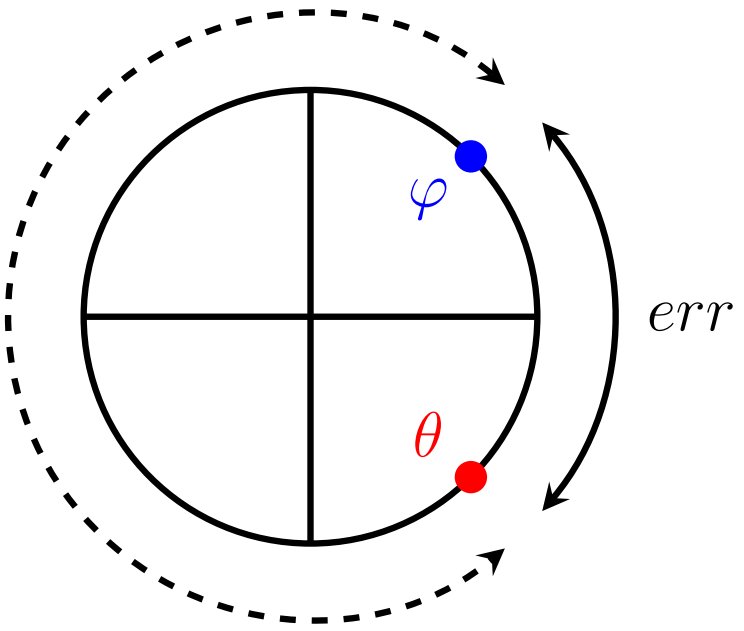

Figure 4: Illustration of *err* on the trigonometric circle: *err* always represent the minimum angular distance between the true $\varphi$ value and an angle $\theta$.

Using the definitions of *err* and $g$ comes

$$\begin{aligned}\mathcal{J}_\varphi &= \int_{-\varphi+\varphi'}^{2\cdot\pi-\varphi+\varphi'} err_{\varphi'}^2(t)\cdot g_{\varphi'}(t)dt \\ &= \underbrace{\int_{-\varphi+\varphi'}^{0} err_{\varphi'}^2(t)\cdot g_{\varphi'}(t)dt}_{\mathcal{M}_\varphi} + \int_{0}^{2\cdot\pi-\varphi+\varphi'} err_{\varphi'}^2(t)\cdot g_{\varphi'}(t)dt\end{aligned} \tag{61}$$

Using the change of variable $u = t + 2\cdot\pi$, and since *err* and $g$ are $2\cdot\pi$-periodic,

$$\mathcal{M}_\varphi = \int_{2\cdot\pi-\varphi+\varphi'}^{2\cdot\pi} err_{\varphi'}^2(u-2\cdot\pi)\cdot g_{\varphi'}(u-2\cdot\pi)du = \int_{2\cdot\pi-\varphi+\varphi'}^{2\cdot\pi} err_{\varphi'}^2(u)\cdot g_{\varphi'}(u)du \tag{62}$$

Leading to

$$\begin{aligned}\mathcal{J}_\varphi &= \int_{0}^{2\cdot\pi-\varphi+\varphi'} err_{\varphi'}^2(t)\cdot g_{\varphi'}(t)dt + \int_{2\cdot\pi-\varphi+\varphi'}^{2\cdot\pi} err_{\varphi'}^2(u)\cdot g_{\varphi'}(u)du \\ &= \int_{0}^{2\cdot\pi} err_{\varphi'}^2(t)\cdot g_{\varphi'}(t)dt = \mathcal{J}_{\varphi'}\end{aligned} \tag{63}$$

The RMSE is thus independent of $\varphi$, and taking $\varphi = 0$ leads to $err(\theta) = |\theta|$ $\forall\theta \in [-\pi, \pi]$, yielding finally

$$\boxed{\text{RMSE} = \sqrt{\int_{-\pi}^{+\pi} \theta^2\cdot g_0(\theta)d\theta}} \tag{64}$$

However, due to the complexity of $g_0(\theta)$, we did not manage to derive a closed-form expression of the RMSE in the general case, and used numerical simulations to compute the latter as a function of $\sigma_x$—or, equivalently, the SNR—$\sigma_p$, and $N$.

The bias, on the other hand, can be readily computed since it can be shown similarly that:

$$\mathbf{E}(\widehat{\varphi} - \varphi) = \int_{-\pi}^{+\pi} \theta \cdot g_0(\theta) d\theta \tag{65}$$

and since $[\theta \mapsto \theta \cdot g_0(\theta)]$ is an odd function, it follows that the bias is null and that $\widehat{\varphi}$ **is an unbiased estimator of** $\varphi$, which further justifies its choice as $\varphi$ estimator in the first place.

### *4.3 Numerical Calculation of the RMSE*

Despite giving a better understanding of the issue at hand, and allowing one to clearly visualize the influence of $N$, $\sigma_p$ or $\sigma_x$ on the probability density function of $\widehat{\varphi}$—as demonstrated by Figures 2 and 3—the polar coordinates considerations did not give a closed form expression for the RMSE of $\widehat{\varphi}$. The latter may thus be calculated numerically using either Equation 46 (Cartesian case) or Equation 64 (polar case):

$$\text{RMSE} = \sqrt{\iint_{\mathbb{R}^2} \arg\left(x + i \cdot y\right)^2 \cdot f(x, y)\ dxdy} = \sqrt{\int_{-\pi}^{+\pi} \theta^2 \cdot g_0(\theta) d\theta} \tag{66}$$

While these two expression of the RMSE obviously yield the same quantity, the Cartesian approach is much more computationally expensive than the polar one, due to the double integration over the discretised, truncated $\mathbb{R}^2$ plane. Thus, all the results below were obtained using the polar approach. The influence of both the $\text{SNR}_{\text{dB}}$—and thus $\sigma_x$—and the phase noise $\sigma_p$ on the RMSE is depicted in Figure 5. Simulated RMSE values were also added to the latter figure, computed in a similar fashion to Figure 3. The closeness between these simulated values and the theoretical predictions consolidates the above-presented approach.

The evolution of the RMSE as a function of the $\text{SNR}_{\text{dB}}$ can be split into three behaviours:

- At very low $\text{SNR}_{\text{dB}}$ values—*i.e.* below −30 dB—the RMSE saturates, to reach approximately 100°. This corresponds to an exceedingly noisy case, for which $\widehat{\varphi}$ is basically no more than a random guess on $[0; 2 \cdot \pi[$. In this case, $g_\varphi(\theta)$ converges towards a uniform distribution, and the RMSE tends towards $\pi/\sqrt{3}$ rad (≈104°), as shown in Section 5.1.

- At higher $\text{SNR}_{\text{dB}}$ value and in the absence of phase noise—*i.e.* when $\text{SNR}_{\text{dB}} > -20$ dB and $\sigma_p$=0—the RMSE follows a linear relationship with the SNR (or a log-linear relationship with SNR$_{\mathrm{dB}}$, as indicates the equation on the graphic). In this case, $g_\varphi(\theta)$ converges towards a normal distribution centred around $\varphi$, and the RMSE tends towards $1/\sqrt{N{\cdot}\mathrm{SNR}}$, as demonstrated in Section 5.2.

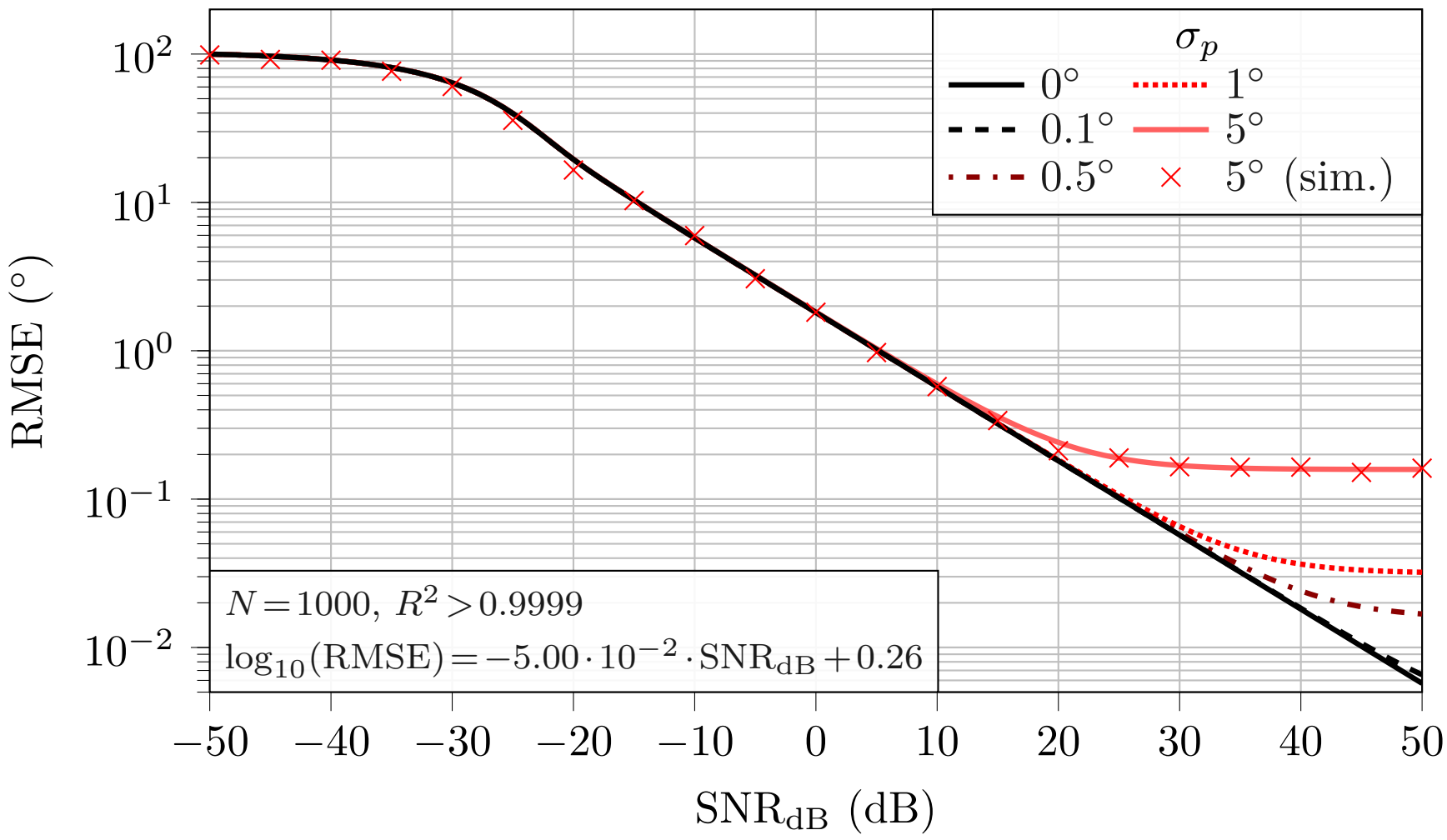

Figure 5: The RMSE on $\varphi$ estimation as a function of SNR$_{\mathrm{dB}}$ for various $\sigma_p$ values. In the absence of phase noise—*i.e.* the black, continuous line (—)—a linear regression can be performed for SNR$_{\mathrm{dB}} \geq -20$ dB, yielding the displayed equation and associated determination coefficient. As indicated, all calculations were performed taking $N$=1000 points. Simulated values were computed using only $10^2$ realisations so that some dispersion is perceivable. Indeed, using $10^6$ realisations as was done in Figure 3 led to marks indistinguishable from the plain red line.

– Finally, when SNR$_{\mathrm{dB}} \geq -20$ dB but the phase noise is significant—typically above 0.5° in Figure 5—the RMSE becomes independent of SNR$_{\mathrm{dB}}$ at some point as SNR$_{\mathrm{dB}}$ increases, and becomes a function of $\sigma_p$ alone. This latter case is developed in Section 5.3.

The influence of $N$ on $\widehat{\varphi}$ RMSE is presented in Figure 6. Starting with the left part of the figure, while the three above-mentioned behaviours can be observed irrespective of $N$ value, increasing $N$ by a factor $m$ has two main effects. First, it shifts the saturation threshold in case of extreme noise—*i.e.* for SNR$_{\mathrm{dB}}$ values in the [-50,-20] dB range—by a factor $10{\cdot}\log_{10}(m)$ (in dB) to the left. Then, it divides the RMSE at higher SNRs—*i.e.* for SNR$_{\mathrm{dB}}$ values in the [-20,50] dB range—by a factor $\sqrt{m}$. Looking at the right part of Figure 6, $\widehat{\varphi}$ RMSE appears to decrease linearly with increasing $N$ values, while increasing the SNR shifts the RMSE lines downwards.

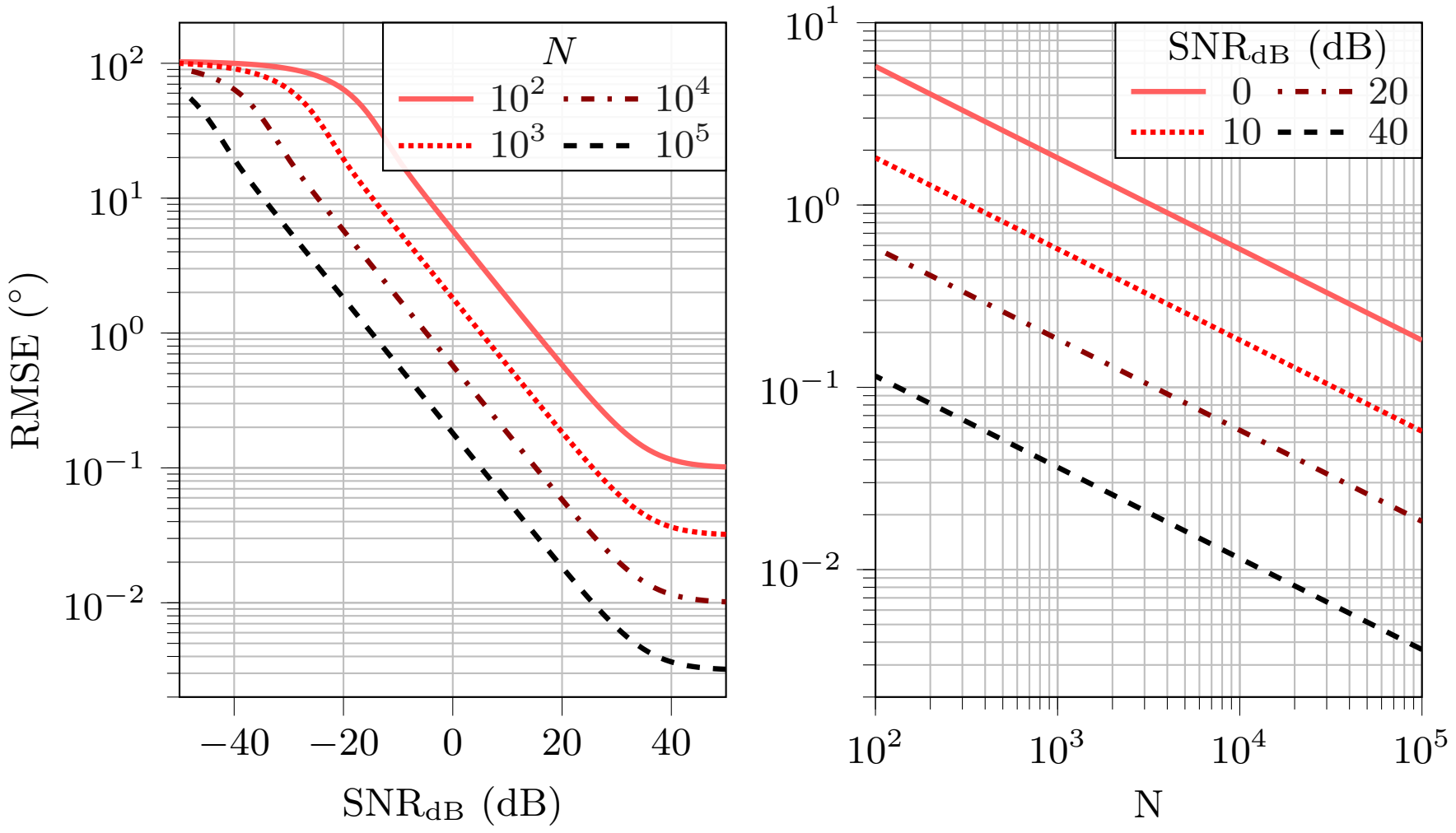

Figure 6: The RMSE on $\varphi$ estimation as a function of $\mathrm{SNR_{dB}}$ and $N$, with $\sigma_p$=1°.

Of note, the reader should bear in mind that the $\mathrm{SNR_{dB}}$ thresholds given above, as well as the numerical values given below, are of course dependent on $N$ and $A_s$. As a general rule, $A_s$ is considered unitary in all our simulations, and $N$=1000 unless otherwise stated. Still, the same above-described behaviours would be observed with different $N$ and $A_s$ values, only the numerical values stated in our developments would be altered. Of note, $\sigma_p$ values were also deliberately chosen relatively high for the sake of illustration. In practice, modern analogue-to-digital converters can feature phase noises in the $[10^{-2}$–$10^{-3}]°$ range[29, 30].

Finally, Figure 7 depicts the evolution of $\widehat{\varphi}$ RMSE with $\sigma_p$. Here, the behaviour observed in Figure 5 can be examined from another perspective: at low enough phase noise levels—*i.e.* when $\sigma_p$ is below approximately 0.01°—the RMSE is solely function of $N$ and the SNR. However, as soon as $\sigma_p$ increases, it starts to act as a noise floor, and increases the RMSE progressively. This effect is particularly pronounced at relatively high $N$ and SNR values, potentially ruining an otherwise excellent accuracy, as can be seen in the left part of the figure for $N = 10^5$ and $\mathrm{SNR_{dB}} = 50$ dB.

### *4.4 Estimation Efficiency*

The efficiency of the performed estimation can be found by computing the Cramér-Rao Lower Bound (CRLB) of $\widehat{\varphi}$. The likelihood $\mathcal{L}\left(\widehat{\varphi}|\widetilde{D}_N\right)$ of observing $\widetilde{D}_N$ under a given $\widehat{\varphi}$—hereafter noted $\mathcal{L}$ for the sake of conciseness—

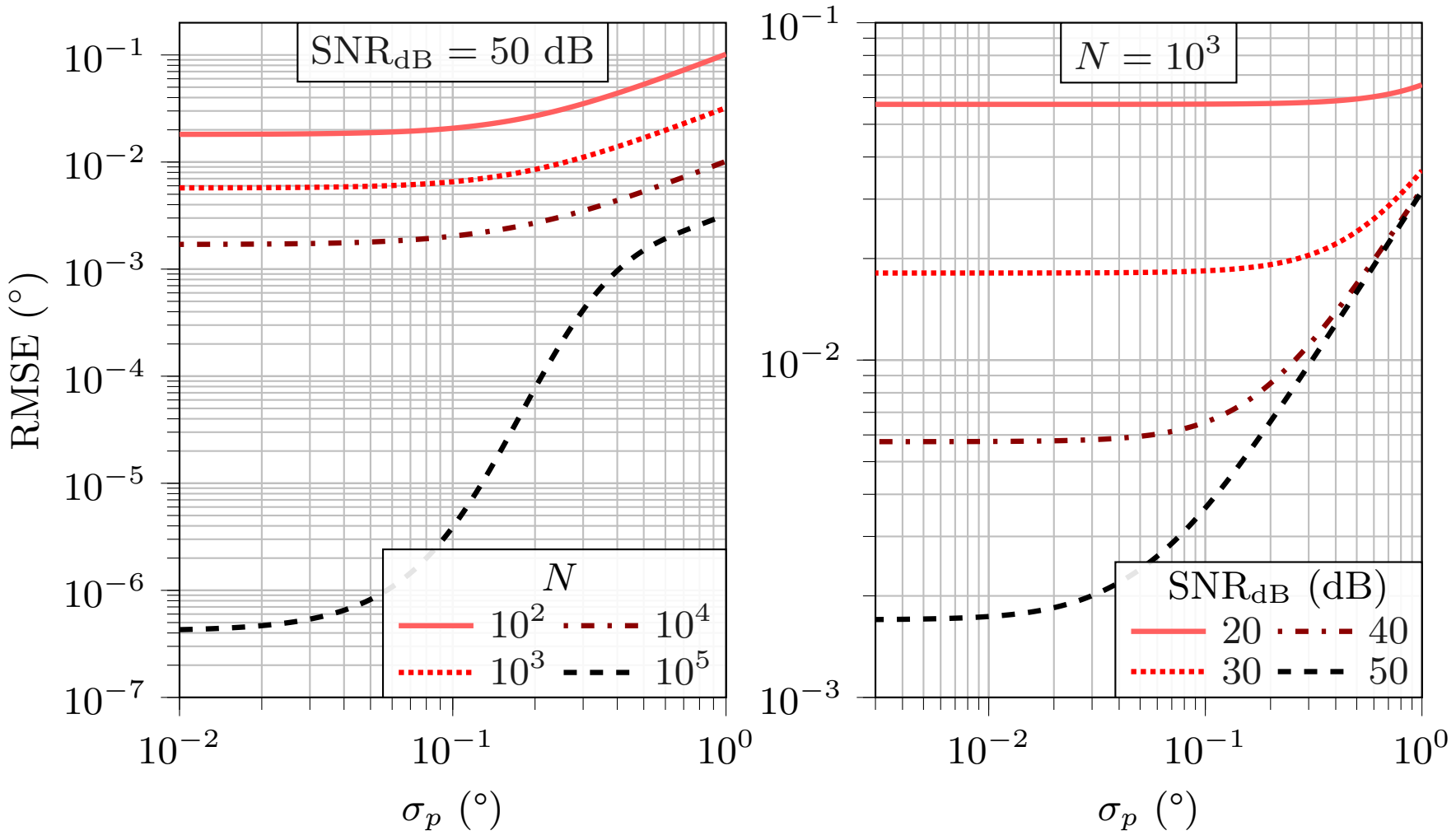

Figure 7: The RMSE on $\varphi$ estimation as a function of $\sigma_p$, for various $N$ and SNR values.

is given by

$$\mathcal{L} = \prod_{z=\{x,y\}} \frac{1}{\sqrt{2{\cdot}\pi{\cdot}\sigma^2}}{\cdot}\mathrm{e}^{-\frac{(z-\mu_z)^2}{2\cdot\sigma^2}} \tag{67}$$

The log-likelihood is then

$$\log\left(\mathcal{L}\right) = -\log\left(2{\cdot}\pi{\cdot}\sigma^2\right) - \sum_{z=\{x,y\}} \frac{\left(z-\mu_z\right)^2}{2{\cdot}\sigma^2} \tag{68}$$

and its first and second derivatives are given by

$$\frac{\partial \log\left(\mathcal{L}\right)}{\partial\widehat{\varphi}} = \frac{1}{\sigma^2}\sum_{z=\{x,y\}} \left(z-\mu_z\right)\cdot\frac{\partial\mu_z}{\partial\widehat{\varphi}} \tag{69}$$

and

$$\frac{\partial^2 \log\left(\mathcal{L}\right)}{\partial\widehat{\varphi}^2} = \frac{1}{\sigma^2}\sum_{z=\{x,y\}}\left[\left(z-\mu_z\right)\cdot\frac{\partial^2\mu_z}{\partial\widehat{\varphi}^2} - \left(\frac{\partial\mu_z}{\partial\widehat{\varphi}}\right)^2\right] \tag{70}$$

The Fisher information $\mathcal{I}\left(\widehat{\varphi}\right)$ may then be derived as

$$\begin{aligned}\mathcal{I}\left(\widehat{\varphi}\right) &= -\mathbf{E}\left[\frac{\partial^2\log\left(\mathcal{L}\right)}{\partial\widehat{\varphi}^2}\right] = \frac{1}{\sigma^2}\sum_{z=\{x,y\}}\left[\left(\frac{\partial\mu_z}{\partial\widehat{\varphi}}\right)^2 - \mathbf{E}\underbrace{\left[z-\mu_z\right]}_{=0}\cdot\frac{\partial^2\mu_z}{\partial\widehat{\varphi}^2}\right] \\ &= \frac{\beta_p^2}{\sigma^2}\end{aligned} \tag{71}$$

Finally leading to

$$\boxed{\mathrm{CRLB} = \frac{1}{\mathcal{I}(\widehat{\varphi})} = \frac{\sigma^2}{\beta_p^2}} \tag{72}$$

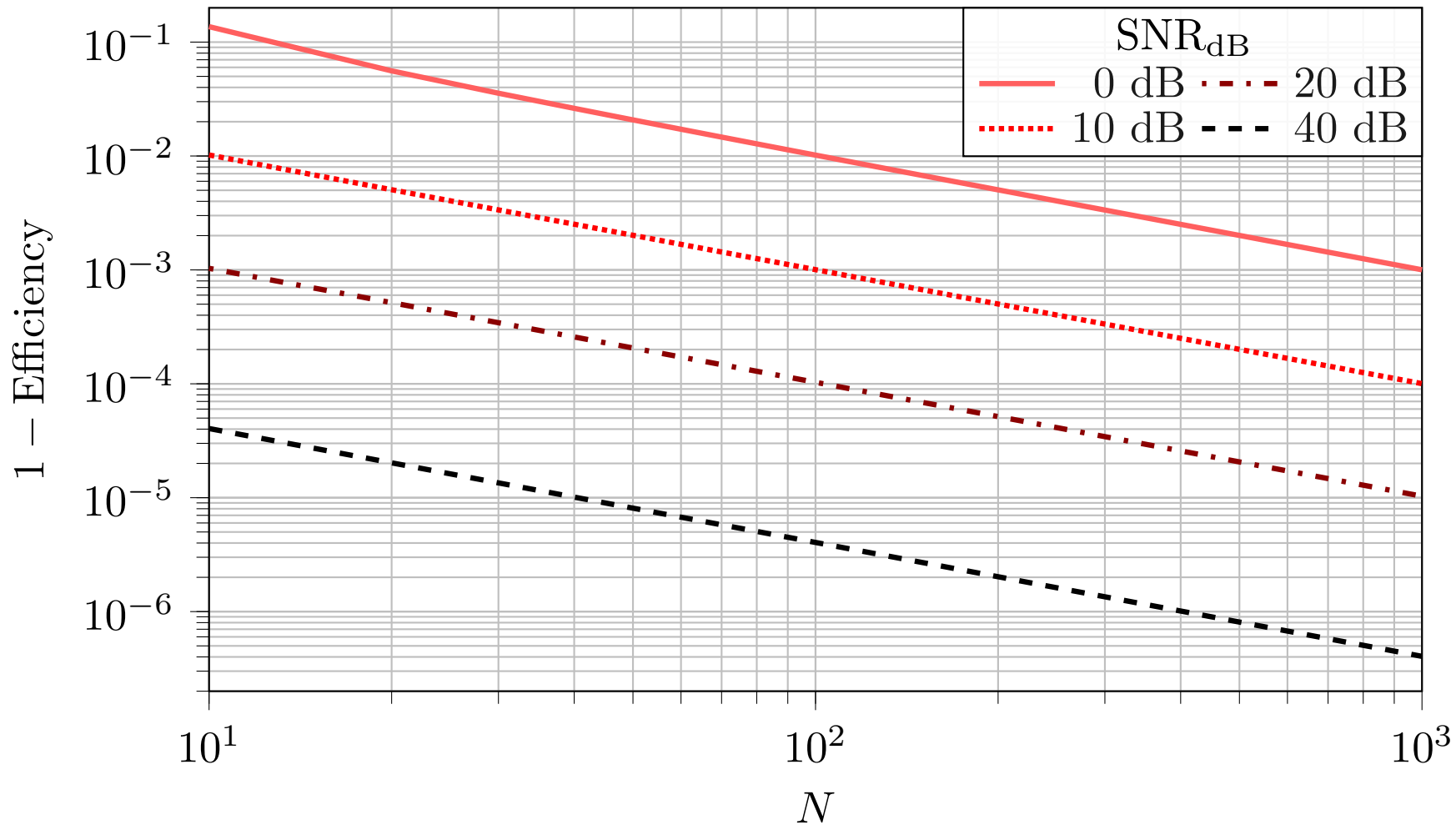

Figure 8: Asymptotical efficiency of $\widehat{\varphi}$. $\sigma_p$ was set to 1°.

The convergence of $\widehat{\varphi}$ towards its CRLB is illustrated in Figure 8. $\widehat{\varphi}$ appears to be an asymptotically efficient estimator of $\varphi$ with a fast convergence rate, exhibiting 1−Efficiency values below $10^{-3}$ for as little as 1000 samples even in the presence of strong noise (SNR=0 dB). As a remainder, the efficiency of an unbiased estimator—which is the case for $\widehat{\varphi}$—is defined as[22, p. 279]

$$\text{Efficiency} = \frac{\mathrm{CRLB}}{\mathbf{Var}(\widehat{\varphi})} = \frac{\mathrm{CRLB}}{\mathrm{RMSE}^2} \tag{73}$$

## 5 Asymptotical Behaviours of the RMSE

### *5.1 Saturation in Case of Excessive Additive Noise*

For SNR$_{\mathrm{dB}}$ values below approximately −30 dB, the RMSE appears to be converging towards approximately 100°—see Figure 5. This phenomenon corresponds to an extremely noisy case, wherein the estimated $\varphi$ value is no better than a random guess on the $[0; 2{\cdot}\pi[$ interval. In this case, $g_0(\theta)$ converges toward a uniform distribution as SNR$_{\mathrm{dB}}$ decreases, as illustrated in Figure 9.

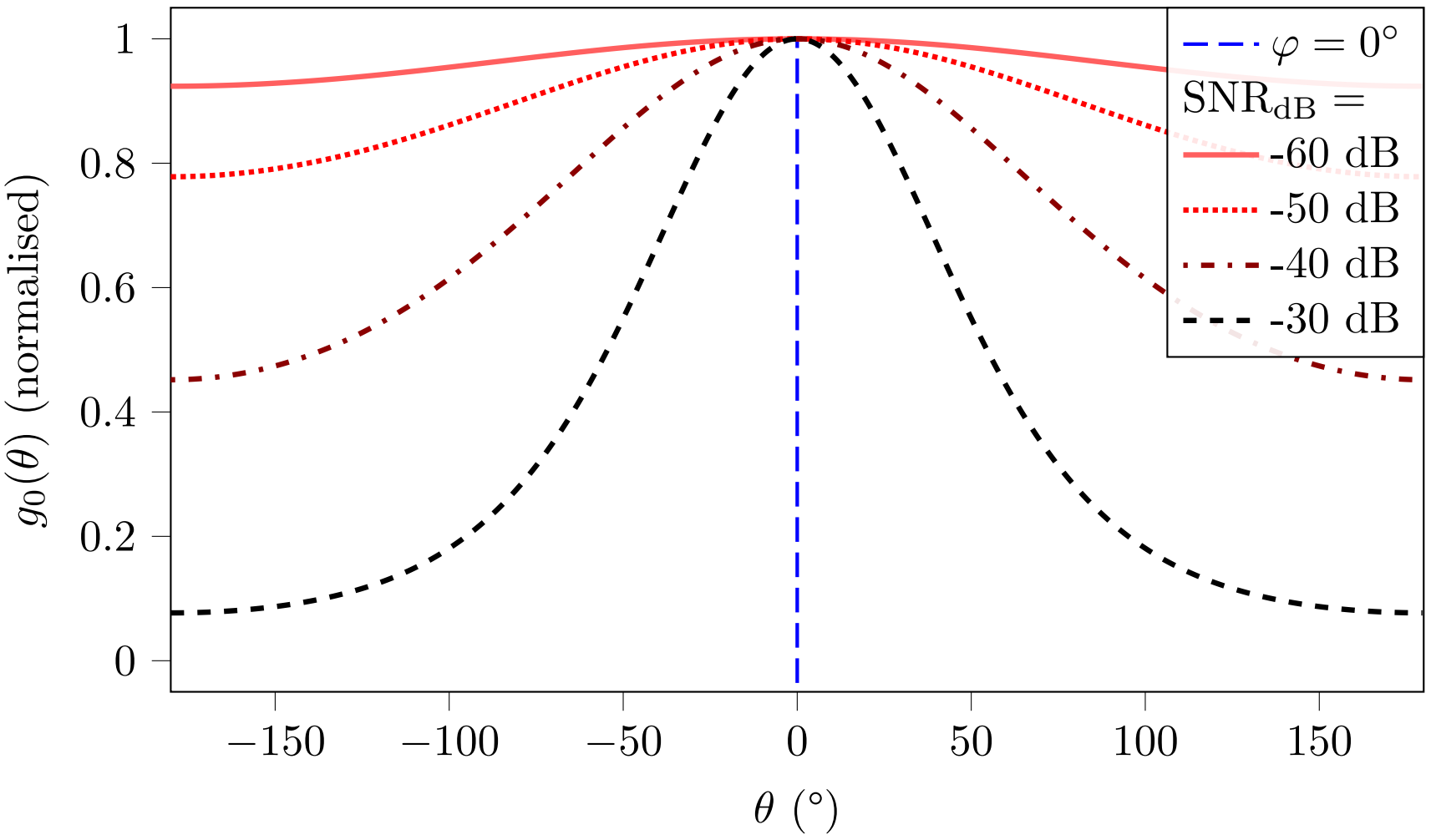

Figure 9: $g_0(\theta)$ for different $\mathrm{SNR_{dB}}$ values. For increasing $\mathrm{SNR_{dB}}$ values, $g_0(\theta)$ converges towards a uniform distribution. $N$=1000, $\sigma_p$=0.

Indeed, as $\mathrm{SNR_{dB}}$ decreases, $\sigma$ increases and we can make the following approximation

$$g_0(\theta) = \frac{1}{2\cdot\pi}\cdot \underbrace{\mathrm{e}^{-\frac{\beta_p^2}{2\sigma^2}}}_{\xrightarrow[\sigma\to+\infty]{}1} + \overbrace{\underbrace{\frac{\overbrace{\beta_p\cdot\cos(\theta)}^{|\ldots|\leq 1}}{2\cdot\sqrt{2\cdot\pi}\cdot\sigma}}_{\xrightarrow[\sigma\to+\infty]{}0}\cdot\underbrace{\mathrm{e}^{-\beta_p^2\cdot\frac{\sin^2(\theta)}{2\cdot\sigma^2}}}_{<1}\cdot\underbrace{\mathrm{erfc}\left(\underbrace{-\beta_p\cdot\frac{\cos(\theta)}{\sigma\cdot\sqrt{2}}}_{\approx 0}\right)}_{\xrightarrow[\sigma\to+\infty]{}1}}^{\xrightarrow[\sigma\to+\infty]{}0} \tag{74}$$

$$g_0(\theta) \xrightarrow[\sigma\to+\infty]{} \frac{1}{2\cdot\pi}$$

In other words, $g_0(\theta)$ can be approximated by a uniform distribution on the $[-\pi,\pi[$ interval. The Kullback-Leibler divergence between $g_0(\theta)$ and a uniform distribution can also be computed, and is presented in Figure 10. As expected, the divergence decreases steeply with a decreasing $\mathrm{SNR_{dB}}$, further confirming the above-mentioned convergence phenomenon.

Subsequently the RMSE becomes

$$\mathrm{RMSE} = \sqrt{\int_{-\pi}^{\pi}\theta^2\cdot\frac{1}{2\cdot\pi}d\theta} = \frac{\pi}{\sqrt{3}} \tag{75}$$

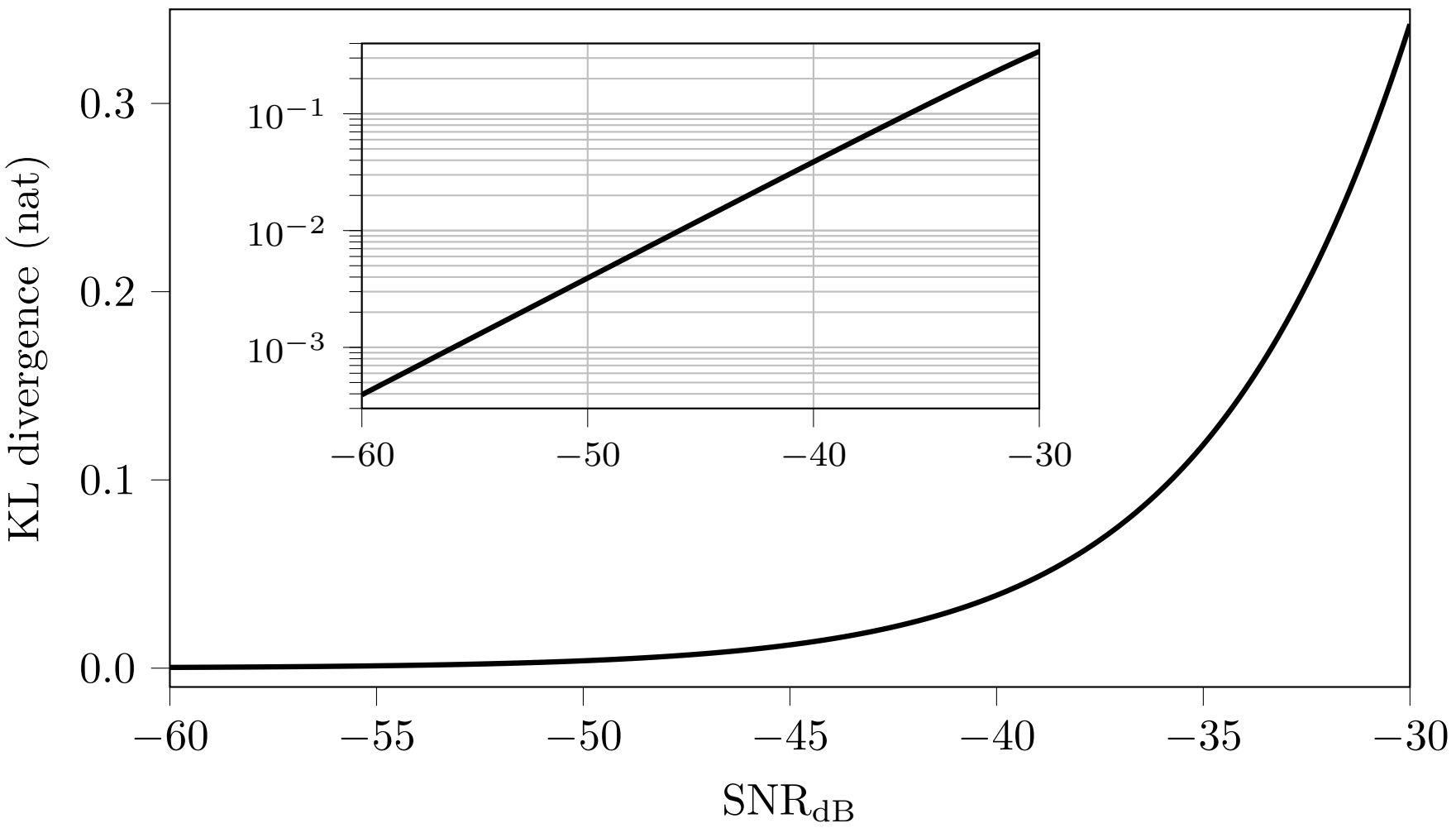

Figure 10: The Kullback-Leibler divergence between $g_0(\theta)$ and a uniform distribution as a function of $\text{SNR}_{\text{dB}}$. The inset shows the same data on a logarithmic scale. $N$=1000, $\sigma_p$=0.

hence the saturation behaviour observed for the RMSE at high $\sigma_c$ values in Figure 5. The 100° plateau value noted above simply comes from the radian to degree conversion ($\pi/\sqrt{3}$ rad≈104°). That being said, contrary to the linear case given in the next section, this saturation phenomenon is of little practical interest since it corresponds to an extremely noisy case, which only yields random guesses as phase estimation. It was thus only presented here for the sake of completeness.

### *5.2 Linear Relationship With the SNR*

The observed linear relationship between the RMSE and the SNR can be explained by the fact that, for increasing SNR values, $g_0(\theta)$ converges towards a normal distribution, as illustrated in Figure 11.

Indeed, taking $\varphi = 0$ and high $\text{SNR}_{\text{dB}}$ values leads to an extremely narrow $g_0(\theta)$ function—as was already noted above in Figure 3—which is non-negligible only for very small $\theta$ deviations from zero. More formally:

$$\begin{aligned} &\mathbf{Var}(\widetilde{D}_N) \xrightarrow[\text{SNR}_{\text{dB}}\to+\infty]{} 0 \quad (*) \\ (*) \implies &\widetilde{D}_N \xrightarrow[\text{SNR}_{\text{dB}}\to+\infty]{L_2} \mathbf{E}[\widetilde{D}_N] \\ (*) \implies &\theta = \arg(\widetilde{D}_N) - \varphi \xrightarrow[\text{SNR}_{\text{dB}}\to+\infty]{P} 0 \end{aligned} \tag{76}$$

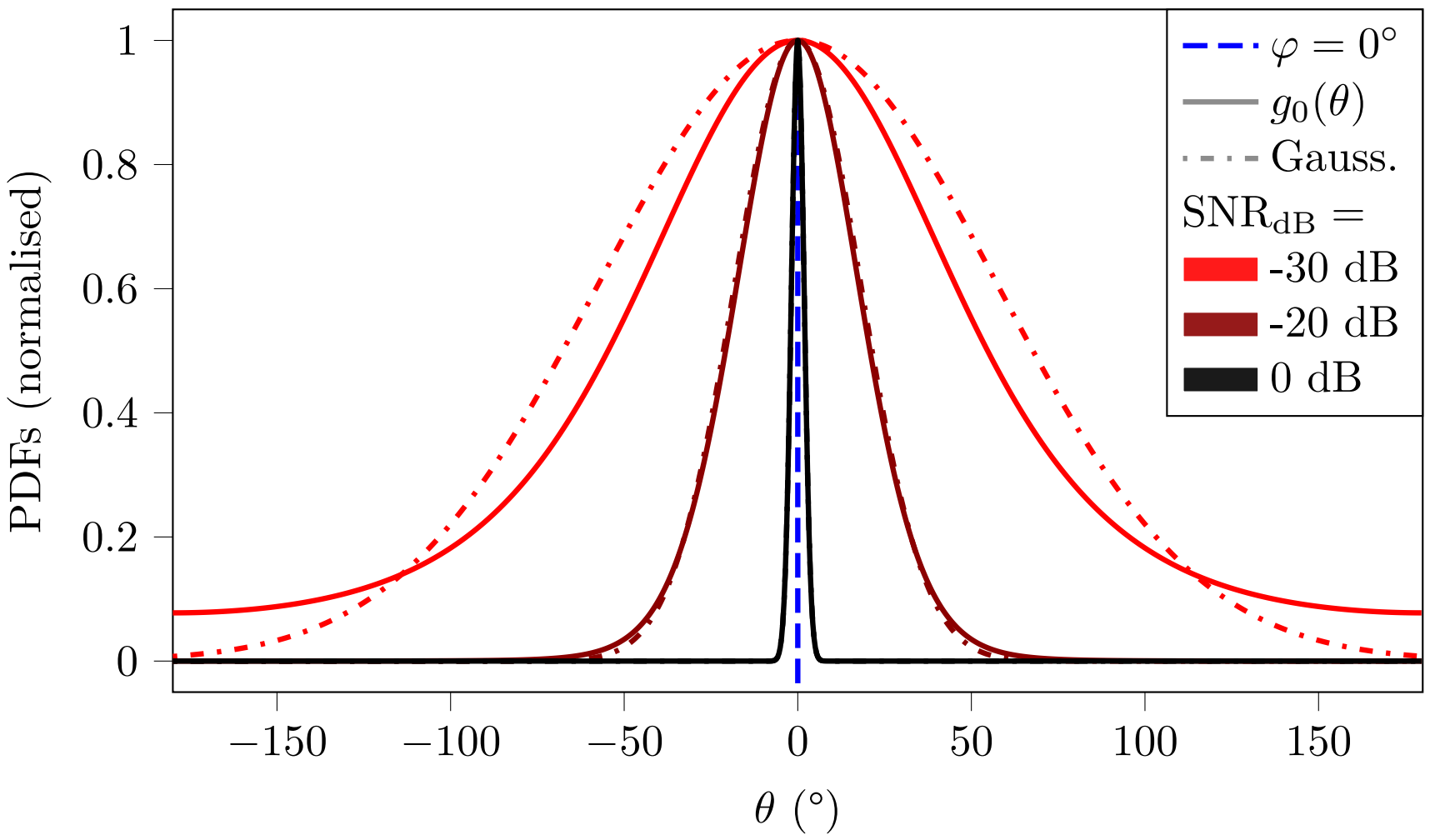

Figure 11: Normalised probability density functions: $g_0(\theta)$ is plotted with its associated simplified Gaussian model. For $\mathrm{SNR_{dB}}$ above $-10$ dB, $g_0(\theta)$ becomes indistinguishable from its Gaussian approximation. $N$=1000, $\sigma_p$=0.

wherein $\xrightarrow{L_2}$ and $\xrightarrow{P}$ denote convergence in mean square and convergence in probability, respectively. By definition of the latter convergence, $g_0(\theta)$ is thus non-negligible only for very small deviations from zero with a non-null probability. Under these conditions, and in the absence of phase noise, $\beta_p = 1$ and $\sigma$ tends towards zero as $\mathrm{SNR_{dB}}$ tends towards infinity. We can then make the following approximation:

$$g_0(\theta) = \underbrace{\frac{\mathrm{e}^{-\frac{\beta_p^2}{2\sigma^2}}}{2\cdot\pi}}_{\approx 0} + \frac{\beta_p}{2\cdot\sqrt{2\cdot\pi}\cdot\sigma}\cdot\underbrace{\cos(\theta)}_{\approx 1}\cdot\mathrm{e}^{-\beta_p^2\cdot\frac{\overbrace{\sin^2(\theta)}^{\approx\theta^2}}{2\cdot\sigma^2}}\cdot\underbrace{\mathrm{erfc}\left(\underbrace{-\beta_p\cdot\frac{\overbrace{\cos(\theta)}^{\approx 1}}{\sigma\cdot\sqrt{2}}}_{\ll -1}\right)}_{\approx 2} \tag{77}$$

$$\approx \frac{1}{\sqrt{2\cdot\pi}\cdot\sigma}\cdot\mathrm{e}^{-\frac{\theta^2}{2\cdot\sigma^2}}$$

Thus, for high enough SNR values, $g_0(\theta)$ can be reasonably well approximated by a simple normal distribution of null mean, and variance $\sigma^2$. The fast convergence towards this approximation is further illustrated in Figure 12, wherein the Kullback-Leibler divergence and Bhattacharyya distance between $g_0(\theta)$ and its Gaussian approximation are represented as a function of $\mathrm{SNR_{dB}}$. It appears that for $\mathrm{SNR_{dB}}$ values above $-10$ dB, this distance becomes virtually null ($\approx 10^{-5}$).

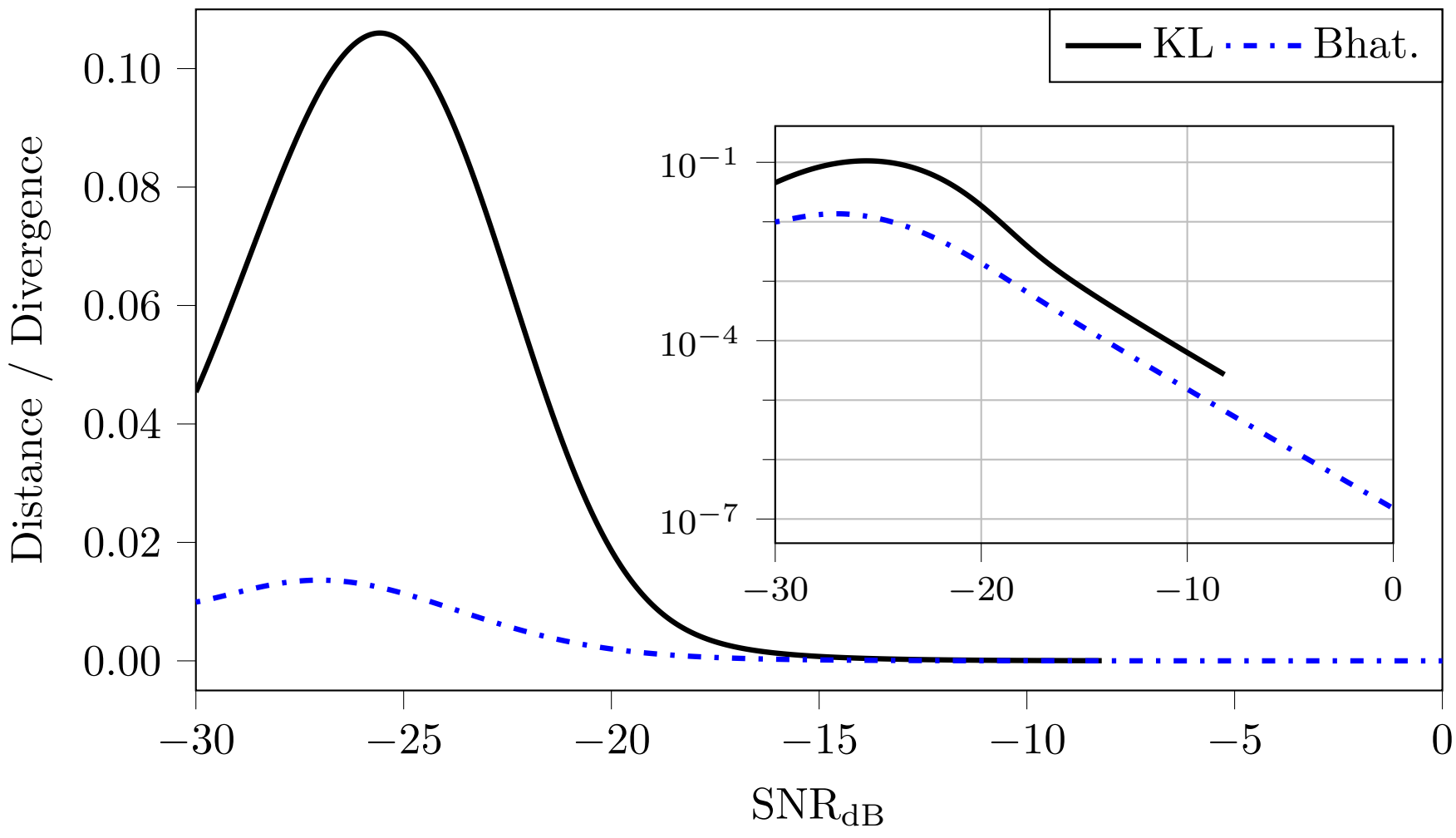

Figure 12: The Kullback-Leibler divergence (KL) and Bhattacharyya distance (Bhat.) between $g_0(\theta)$ and its Gaussian approximation as a function of $\mathrm{SNR_{dB}}$. The inset shows the same data on a logarithmic scale. $N$=1000, $\sigma_p$=0.

As a reminder, the Bhattacharya distance quantifies the *closeness* between two distributions. Given two random variables $P$ and $Q$ with probability density function $p(x)$ and $q(x)$ the Bhattacharya distance and Kullback-Leibler divergence are defined as follows[31]:

*(i)* Bhattacharya distance:

$$d_B(P\|Q) = -\log\left(\int p(x)\cdot q(x)\;dx\right) \tag{78}$$

*(ii)* Kullback-Leibler divergence:

$$d_{KL}(P\|Q) = \int p(x)\cdot\log\left(\frac{p(x)}{q(x)}\right)dx \tag{79}$$

One may notice that the division by $q(x)$ in $d_{KL}$ may be problematic in case of numerical applications if $q(x)$ is near zero. Such an issue can be observed in Figure 12 for $\mathrm{SNR_{dB}}$ values above roughly $-8$ dB: $d_{KL}$ cannot be computed even using 64-bits double-precision floats, because the Gaussian probability density function tends towards zero extremely fast as soon as $\theta$ deviates from zero. This is the reason why the Bhattacharya distance was introduced in the first place, so as to better cover the case of high $\mathrm{SNR_{dB}}$ values.

Subsequently the RMSE becomes

$$\mathrm{RMSE} = \sqrt{\int_{-\pi}^{\pi} \theta^2 \cdot \frac{1}{\sqrt{2\cdot\pi}\cdot\sigma} \cdot \mathrm{e}^{-\frac{\theta^2}{2\cdot\sigma^2}} d\theta} = \sigma = \frac{1}{\sqrt{N\cdot\mathrm{SNR}}} \tag{80}$$

hence the linear relationship observed in Figure 6 between $\log_{10}(\mathrm{RMSE})$ on the one hand, and $\mathrm{SNR}_{\mathrm{dB}}$ and $N$ on the other hand. This result is especially interesting for practical applications. Indeed, a $\mathrm{SNR}_{\mathrm{dB}}$ above $-10$ dB can easily be reached in practice, ensuring the validity of the above-mentioned Gaussian approximation for $g_0(\theta)$, and the ensuing conclusions on the RMSE. Of particular interest, if this condition is fulfilled, **the expected RMSE on the phase estimation can be directly inferred from N and the SNR**, using Equation 80. Still, one should bear in mind that this conclusions holds only if $\beta_p \approx 1$, otherwise the developments presented in the next section should apply.

### *5.3 Saturation in the Presence of Phase Noise*

When some amount of phase noise is present—*i.e.* $\beta_p \in ]0;1[$—and for high enough SNR, the following approximation can be made, as was done in Equation 77:

$$g_0(\theta) \approx \frac{\beta_p}{\sqrt{2\cdot\pi}\cdot\sigma} \cdot \mathrm{e}^{-\frac{\beta_p^2\cdot\theta^2}{2\cdot\sigma^2}} \tag{81}$$

Again, $g_0(\theta)$ can be approximated by a simple normal distribution, but of variance $\sigma^2/\beta_p^2$, instead of $\sigma^2$ alone in the previous section. The convergence towards a Gaussian is nearly identical with that presented in Section 5.2 and thus Figures 11 and 12 were not reproduced for the sake of conciseness. Similarly to Equation 80 comes

$$\mathrm{RMSE} = \frac{\sigma}{\beta_p} = \sqrt{\frac{1}{\beta_p^2\cdot N} \cdot \left(1 - \beta_p^2 + \frac{1}{\mathrm{SNR}}\right)} \tag{82}$$

which, in case of high enough SNR—*i.e.* $1 - \beta_p^2 \gg 1/\mathrm{SNR}$—becomes

$$\mathrm{RMSE} \approx \sqrt{\frac{1}{N} \cdot \left(\frac{1}{\beta_p^2} - 1\right)} \tag{83}$$

This result is also interesting because it provides a lower limit for the RMSE, even at exceedingly large SNR values: **the RMSE is ultimately limited by the phase noise, which acts as a noise floor**. This explains the saturations observed on the right part of Figure 5 and 6: the lower limits reached by the different curves with non-zero phase noises directly depend on their respective $\sigma_p$ values, following Equation 83. Most interestingly, Equation 82 gives a generic expression for the RMSE at reasonably high SNR values taking into account the joint influences of: *(i)* the number of points $N$, *(ii)* the amplitude of the phase noise—through $\beta_p$—and *(iii)* that of the additive noise—through the SNR.

## 6 Conclusion

This article presents a thorough analysis of the influence of additive and phase noises on the accuracy of the phase measurement of a known-frequency sinusoidal signal. More specifically, we focused on synchronous detection, a measurement scheme for which the number of collected samples $N$ on the one hand, and the sampling and probing frequency $f_s$ and $f_0$ on the other hand can be chosen so that $f_0 \cdot N/f_s$ is an integer. In this particular case, a closed-form expression of the PDF of the phase estimate $\widehat{\varphi}$ could be derived, depending on $N$ and on the levels of additive and phase noises—$\sigma_x$ and $\sigma_p$, respectively. $\widehat{\varphi}$ was also shown to be asymptotically efficient, with a fast convergence towards its CRLB, even using a limited number of samples in the presence of substantial noise levels.

When using the above-mentioned PDF to compute $\widehat{\varphi}$ RMSE, three main behaviours could be identified: *(i)* in case of excessive noise, the RMSE saturates towards the random guess situation, *(ii)* as the SNR increases, the RMSE decreases linearly with the square roots of the SNR and N, and *(iii)* as the SNR further decreases, the RMSE saturates again, reaching a noise floor caused by phase noise. While *(i)* is of little practical interest, *(ii)* and *(iii)* are of major importance in practical scenarios, since they can tell the experimenter whether lowering the RMSE should be achieved by taking more samples or by increasing the SNR—*e.g.* by increasing the emission power. Indeed, in case *(iii)*, increasing the SNR is no use once the phase noise floor is reached, and only taking more samples can yield lower RMSE values.

In addition to its theoretical significance, this paper is thus also of practical value, allowing for informed decision-making when designing a phase-measuring apparatus. In particular, in the context of f-DLR mentioned in introduction, a compromise often has to be made between reducing the number of samples and the illumination power—which preserves the involved dyes from photobleaching, and saves power in case of battery-powered devices—and reducing the RMSE by increasing the two latter parameters—at the expense of power consumption, computing costs, and dye photobleaching. Of particular importance, we recently published a short communication demonstrating a real-world implementation of f-DLR applied to Carbon dioxide ($CO_2$) sensing[32]. In the latter work, we validated the theoretical results derived in the present paper in the case of a low phase noise and a relatively high SNR—*i.e.*

Equation 80.

Finally, one possible limitation of this work resides in its simplified modelling of phase noise. Indeed, phase noise—studied through its influence on the power spectral density of a clock signal in the frequency domain[33], or by the distribution of the jitter it induces in the time domain[34]—can result from several phenomena that can be either purely random or partially deterministic[35, Chap. 2–4]. While a thorough review on the origins and statistical properties of phase noise is clearly out of the scope of this paper, and although some oscillators do exhibit Gaussian phase jitters in practice[36, 37, 38], this may not be the case in general[34]. As a result, future research could focus on extending our work to address non-Gaussian and/or non-*i.i.d.* phase noise.

## Acknowledgments

We are grateful to Yoshitate Takakura for his valuable insights in spectral estimation, and to Morgan Madec for his meticulous early review of this work.